\documentclass[twocolumn,reprint,amsmath,amssymb,footinbib,superscriptaddress,floatfix,aps,longbibliography,prl]{revtex4-2}
\pdfoutput=1

\usepackage{hyperref}
\usepackage[utf8]{inputenc}
\usepackage[sort&compress]{natbib}
\usepackage[dvipsnames]{xcolor}
\usepackage{graphicx}   
\usepackage{epsfig}
\usepackage{epstopdf}
\usepackage{amsfonts}
\usepackage{lipsum}
\usepackage{amstext,amsmath,amssymb}

\usepackage{rotating}
\usepackage{dcolumn}
\usepackage{array,multirow}
\usepackage[english]{babel}
\usepackage[separate-uncertainty=true]{siunitx}

\DeclareSIUnit\stokes{St}
\DeclareSIUnit\gauss{G}
\usepackage[export]{adjustbox}
\usepackage{braket}
\usepackage[caption=false]{subfig}
\usepackage{xr}
\usepackage{ifpdf}
\usepackage{tikz}
\usetikzlibrary{arrows}
\usepackage{natbib}
\usepackage{graphicx}
\usepackage{bm}
\usepackage{amsfonts}
\usepackage{amsmath}

\usepackage{bbold}
\usepackage{amsmath}

\usepackage{orcidlink}
\begin{document}

\title{Protecting Quantum Information via Destructive Interference of Correlated Noise}
\author{Alon Salhov \orcidlink{0000-0001-9111-2400}}
\thanks{These authors contributed equally.\\ \href{mailto:alon.salhov@mail.huji.ac.il}{alon.salhov@mail.huji.ac.il}
}
\affiliation{Racah Institute of Physics, The Hebrew University of Jerusalem, Jerusalem, 91904, Givat Ram, Israel}
\author{Qingyun Cao}
\thanks{These authors contributed equally.\\ \href{mailto:alon.salhov@mail.huji.ac.il}{alon.salhov@mail.huji.ac.il}
}
\affiliation{Institute for Quantum Optics, Ulm University, Albert-Einstein-Allee 11, 89081 Ulm, Germany}
\affiliation{School of Physics, International Joint Laboratory on Quantum Sensing and Quantum Metrology, Huazhong University of Science and Technology, Wuhan 430074, China}
\author{Jianming Cai}
\affiliation{School of Physics, International Joint Laboratory on Quantum Sensing and Quantum Metrology, Huazhong University of Science and Technology, Wuhan 430074, China}
\author{Alex Retzker}
\affiliation{Racah Institute of Physics, The Hebrew University of Jerusalem, Jerusalem, 91904, Givat Ram, Israel}
\affiliation{AWS Center for Quantum Computing, Pasadena, CA 91125, USA}
\author{Fedor Jelezko}
\affiliation{Institute for Quantum Optics, Ulm University, Albert-Einstein-Allee 11, 89081 Ulm, Germany}
\author{Genko Genov \orcidlink{0000-0002-4538-6686}}
\affiliation{Institute for Quantum Optics, Ulm University, Albert-Einstein-Allee 11, 89081 Ulm, Germany}

\date{\today}

\begin{abstract}
Decoherence and imperfect control are crucial challenges for quantum technologies. Common protection strategies rely on noise temporal autocorrelation, which is not optimal if other correlations are present. We develop and demonstrate experimentally a strategy that utilizes the cross-correlation of two noise sources. We achieve a tenfold coherence time extension by destructive interference of cross-correlated noise, improve control fidelity, and surpass the state-of-the-art sensitivity for high frequency quantum sensing, significantly expanding the applicability of noise protection strategies.
\end{abstract}

\maketitle

\paragraph{Introduction.---}\label{sec:Introduction}
Decoherence, typically caused by unwanted couplings to the environment and control noise \cite{ZurekRMP2003,khodjasteh2005fault}, remains a major challenge for quantum technologies. Quantum computation requires reducing its effect to achieve the long memory time and high gate fidelity required for fault-tolerance \cite{aharonov1997fault,shor1996fault}. The sensitivity of quantum sensors typically scales with the sensor's coherence time and is thus also limited by decoherence \cite{Degen2017RMP}. 
While fabrication efforts focus on minimizing noise in quantum devices \cite{siddiqi2021engineering} and quantum error correction techniques allow detection and correction of noise-induced errors \cite{terhal2015quantum,lidar2013quantum}, several important strategies, such as decoherence-free spaces \cite{lidar2003decoherence}, clock transitions \cite{wolfowicz2013atomic}, dynamical decoupling \cite{Viola1999PRL} and composite pulses \cite{Levitt1986,Wimperis1994,Jones2003PRA,Torosov2011,Genov2013PRL,Genov2014,Hain2020pra,Torosov2022PRL}, reduce the effect of noise, lowering decoherence and control error rates \cite{Suter2016RevModPhys}. 

Each strategy takes advantage of a ``resource'' to protect quantum information. Decoherence-free-subspaces, for example, employ symmetries in system-bath coupling by storing quantum information in subspaces with low noise susceptibility. Temporal auto-correlations of noise constitute another resource, which dynamical decoupling and composite pulses utilize to partially refocus the effect of system-environment interactions and control noise. 

In this work, we propose and experimentally demonstrate a protection strategy that relies on a different kind of resource -- the cross-correlation of two noise sources, e.g. control fields.
Such cross-correlations exist when the control fields are generated from the same source or pass through the same transmission line. As an example, we modify the continuous concatenated dynamical decoupling control scheme \cite{CaiNJP2012}, which has been experimentally demonstrated for coherence protection and quantum sensing \cite{StarkNatComm2017,Joas2017NatComm,wang2020coherence,https://doi.org/10.48550/arxiv.2207.06611,ramsay2023coherence,wang2021observationMollow,CohenFP2017,farfurnik2017experimental,teissier2017hybrid,cao2020protecting,Genov2019MDD}.
As we show, introducing a frequency shift to one of the control fields, which is proportional to the degree of cross-correlation, results in destructive interference of the cross-correlated noise. Our scheme gives an order-of-magnitude enhancement of coherence time, compared to the standard technique, and is limited mainly by the lifetime of the qubit. 
We use it for improved quantum sensing and robust qubit operations, demonstrating its advantages and broad applicability.

\paragraph{Theory.---}\label{sec:Theory}

We consider a two-level system with a Hamiltonian ($\hbar=1$) 
\begin{equation}\label{Eq:lab_frame_H_MT}
    \begin{split}
       H&=
       \frac{1}{2}(\omega_0+\delta(t))\sigma_{z} 
       +\Omega_1(1+\epsilon_1(t)) \cos{(\omega_0 t)}\sigma_{x} \\
       &
       -2\Omega_2 (1+\epsilon_2(t))\sin{\left(\omega_0 t
       \right)}\cos{(\widetilde{\Omega}_1 t)}\sigma_{x},
    \end{split}
\end{equation}
where $\omega_0$ is the qubit bare energy gap, $\delta(t)$ is an environment-induced noise term, $\Omega_1$ is the Rabi frequency of a protective field, perpendicular to the noise \cite{CaiNJP2012}, with $\epsilon_1(t)$ its relative error \cite{Supplemental}. We compensate the first field noise by a second field with Rabi frequency $\Omega_2$, modulation frequency $\widetilde{\Omega}_1$, and relative error $\epsilon_2(t)$. The noise terms are assumed stationary with equal, sub-unity variance ($\overline{\epsilon_i(t)}=0, \sigma^2\equiv\overline{\epsilon_i(t)^2}\ll1$; overbar indicates average over experimental runs). The equal variance assumption is experimentally motivated \cite{kong2018nanoscale,Supplemental}, but can be relaxed. 

\begin{figure}[t]
\centering
\includegraphics[width=0.95\linewidth]{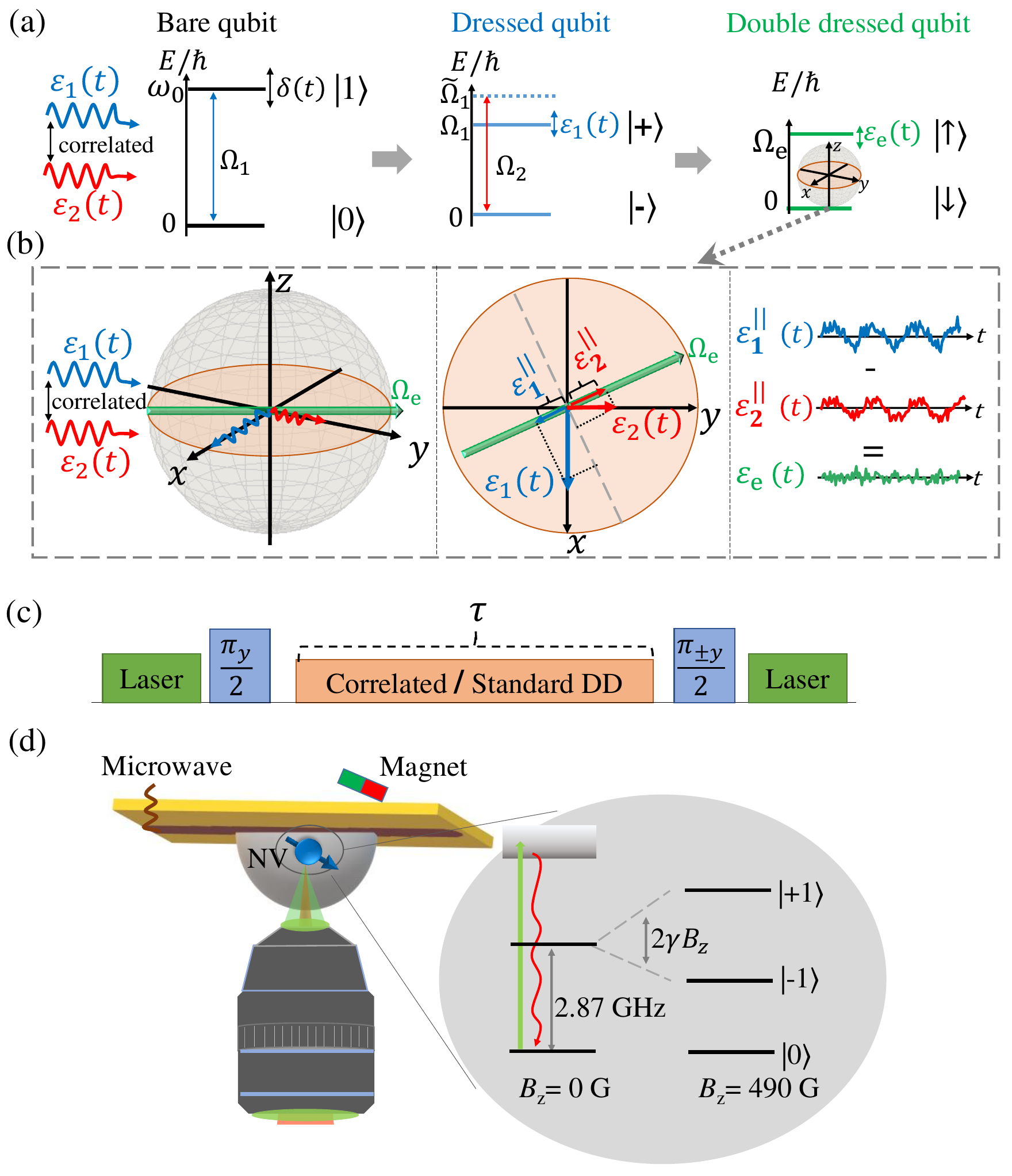}
\centering
\caption{Schematic representation of destructive interference of cross-correlated noise, control sequences and experimental setup. (a) The qubit is subjected to environmental noise $\delta(t)$. Applying a resonant drive with Rabi frequency $\Omega_1$ creates a protected dressed qubit which decoheres mainly due to $\epsilon_1(t)$ - the amplitude noise in $\Omega_1$. Applying a second drive with modulation frequency $\widetilde{\Omega}_1$, Rabi frequency $\Omega_2$ and amplitude fluctuations $\epsilon_2(t)$, reduces decoherence due to $\epsilon_1(t)$. (b) If the cross-correlation, $c$, of $\epsilon_1(t)$ and $\epsilon_2(t)$ is nonzero, a detuning $\widetilde{\Omega}_1=\Omega_1+c\,\Omega_2^2/\Omega_1$ tilts the effective-drive axis and induces a destructive interference of the cross-correlated noise, resulting in a doubly-dressed qubit with a longer coherence time. (c) Measurement sequences for standard and correlated double drive (DD) protocols. (d) Experimental setup and level scheme of the NV center.}
\label{setup}
\end{figure}

The standard double-drive (DD) technique \cite{CaiNJP2012,StarkNatComm2017} relies on the mechanism of continuous dynamical decoupling (see Fig. \ref{setup}(a) and \cite{Supplemental}). Disregarding potential cross-correlations, the noise of the first field $\epsilon_1(t)$ is optimally decoupled at the standard resonance condition $\widetilde{\Omega}_1=\Omega_1$, but the system suffers from the second field noise $\epsilon_2(t)$ \cite{CaiNJP2012,StarkNatComm2017,Supplemental}.

Detuning $\widetilde{\Omega}_1$ from resonance reintroduces the effect of $\epsilon_1(t)$ when there is no cross-correlation between the noise terms, 
i.e., $\overline{\epsilon_1(t)\epsilon_2(t)}=0$. However, non-zero cross-correlations are expected when the fields share control hardware. Then, reintroduction of $\epsilon_1(t)$ is beneficial, if it is set to destructively interfere with $\epsilon_2(t)$. To show this, we transform $H$ into a doubly-rotating-frame at $\omega_0 \sigma_{z}/2$ and then $ \widetilde{\Omega}_1\sigma_{x}/2$, 
and apply the rotating-wave approximation $\Omega_2\ll\Omega_1,\widetilde{\Omega}_1\ll\omega_0$ to obtain \cite{Supplemental}
\begin{equation}\label{Eq:HII_MT}
    H_{II}=\frac{\left(\Omega_1-\widetilde{\Omega}_1\right)+\Omega_1\epsilon_1(t)}{2}\sigma_{x}+\frac{\Omega_2}{2}(1+\epsilon_2(t))\sigma_{y}.
\end{equation}
To prolong the doubly-dressed \cite{cohen1994atoms} qubit's coherence time, we choose $\widetilde{\Omega}_1$ to minimize the variance of its energy gap. The resulting detuning $(\widetilde{\Omega}_1-\Omega_1)$ tilts the effective-drive axis to a correlation-dependent angle. Then, the projections of the correlated noise terms on this axis, which have the first-order effect on decoherence, destructively interfere (see Fig. \ref{setup}(b) and \cite{Supplemental} for analysis and further discussion on cross-spectral densities). The optimal modulation frequency of the second drive, to leading order in $\Omega_2\ll\Omega_1$, reads
\begin{equation}\label{corr_omega1_MT}
\small
    \begin{aligned}
    \widetilde{\Omega}_1
    &
    \approx
    \Omega _1+c\frac{\Omega_2^2}{\Omega_1},~\text{where}~c\equiv\frac{\overline{\epsilon_1(t)\epsilon_2(t)}}{\sigma^2}\\
    \end{aligned}
\end{equation}
is the cross-correlation of the fields' fluctuations. Note that this correlation-induced frequency shift is not related to the Bloch-Siegert shift $\frac{\Omega_2^2}{4\Omega_1}$ \cite{bloch1940magnetic,james2007effective}. It has a different magnitude and physical origin and exists with circularly polarized control fields, where the latter is zero because there are no counter-rotating terms. A complete treatment of both effects requires substituting $c\rightarrow(c+\frac{1}{4})$ into Eq. \eqref{corr_omega1_MT} \cite{Supplemental}.

The \textit{correlated-noise-shift} in Eq. \eqref{corr_omega1_MT} (with the $c\rightarrow(c+\frac{1}{4})$ correction) is the main result of this section and defines the correlated DD protocol. Typically, $c$ is system-dependent, suggesting to scan the detuning and optimize the coherence time \cite{Supplemental}. We observe almost perfect correlation ($c\approx1$) in our experimental setup \cite{Supplemental}, which allows for complete noise suppression to first order as the doubly-dressed states become (dynamic) clock states. The resulting stability of the dressed qubit is better than simply the stability of its components, that is, the bare qubit or the control. 
As Eq. \eqref{corr_omega1_MT} reduces to the standard DD for $c=0$ we conclude that the destructive interference principle is compatible with dynamical decoupling. We distinguish the contribution of the two-field cross-correlation from the standard DD effect by comparing the two protocols. In the following, we demonstrate the superiority of correlated DD for (1) quantum memory, (2) quantum sensing, and (3) robust coherent control.

\paragraph{Quantum Memory.---}
We experimentally demonstrate correlated DD for a quantum memory in a single nitrogen-vacancy (NV) center in diamond. 
%
The diamond sample is produced by chemical vapor deposition (CVD) and polished into a hemisphere, 
acting as a solid immersion lens and enhancing photon collection efficiency \cite{hadden2010strongly,siyushev2010monolithic,Supplemental}. To create NV centers, the surface is overgrown with 100 nm layer of isotopically enriched ${}^{12}\!$C (99.999\%) by plasma-enhanced CVD \cite{osterkamp2015stabilizing}. 
The NV center's negative charge state allows optical detection and polarization of its electron spin \cite{manson2006nitrogen,doherty2013nitrogen}. 
We apply a bias magnetic field of $490$ Gauss parallel to the NV axis to lift the degeneracy of the $m_{s}= \pm 1$ ground states (Fig. \ref{setup}(d)) and polarize the nitrogen nuclear spin 
\cite{jacques2009dynamic}. We use a 532 nm laser to initialize the system in $|0\rangle$. We prepare a superposition state between $|-1\rangle$ and $|0\rangle$ by a microwave $\pi/2$ pulse, apply a control scheme, and then another $\pi/2$ pulse with a phase that alternates between 0 and 180$^{\circ}$, to map coherences back onto populations (Fig. \ref{setup}(c)). 
We estimate the final populations from the difference between the signals, reducing errors due to charge-state and count-rate fluctuations \cite{haberle2017nuclear}. All control fields originate from the same arbitrary waveform generator and amplifier.

\begin{figure}[hpbt]
\centering
\includegraphics[width=1\linewidth]{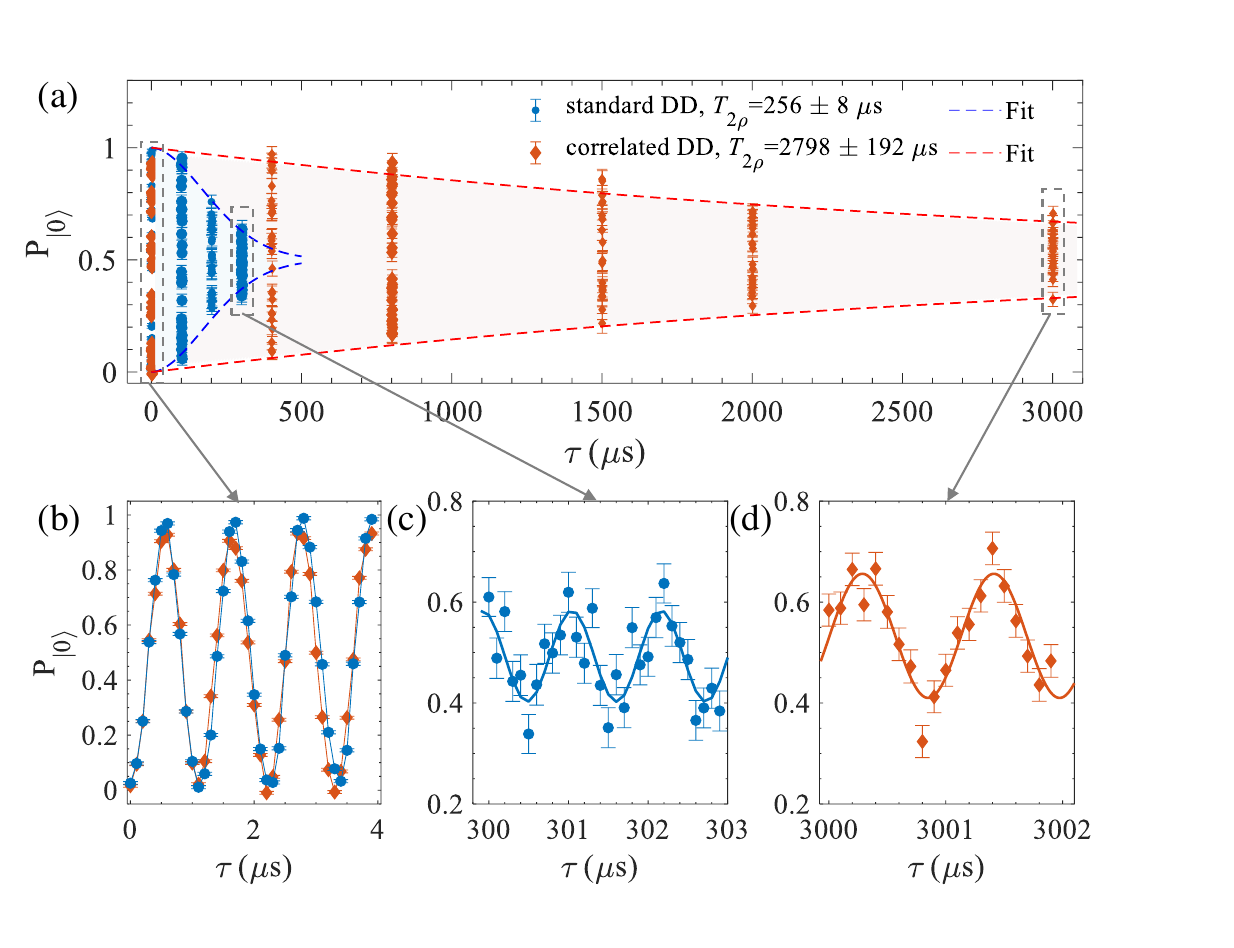}
\centering
\caption{Experimental measurements of the population of state $|0\rangle$ ($P_{|0\rangle}$) vs. experimental time $\tau$, demonstrating improved quantum memory. (a) We fit a stretched exponential function $\exp(-(\tau/T_{2})^\beta)$ to the difference between the higher and lower signal envelopes and obtain $T_{2\rho}\approx 256\pm 8\,\mu s$ ($\beta=1.86\pm 0.17$) for standard DD (blue), which is extended twelvefold to $T_{2\rho}\approx 2.798\pm 0.192$ ms ($\beta=1.04\pm 0.11$) for correlated DD (red). The higher and lower envelope fits are shown with dashed lines. For correlated DD we shift the modulation frequency, $\widetilde{\Omega}_1$, according to Eq. \eqref{corr_omega1_MT} with $c\approx 1$, and the $c \rightarrow c+\frac{1}{4}$ correction. (b), (c), (d) Zoomed-in views of measurements at $1\,\mu$s, $300\,\mu$s and $3$ ms. The frequency of the signal oscillation in the lab frame is $\approx\Omega_2$, as expected from theory.}
	\label{coherence_improvement}
\end{figure}

We first consider a Ramsey measurement \cite{Supplemental} resulting in a coherence time of $T_2^{\ast}\approx 28~\mu$s. 
Decoupling by one continuous field, resonant with the $|0\rangle\leftrightarrow |-1\rangle$ transition, with $\Omega_1=(2\pi)\,4.350$ MHz, increases the coherence time to $T_{2\rho,\text{single-drive}}\approx 110\,\mu$s.
We then use standard DD with $\Omega_1=(2\pi)\,4.470$ MHz and $\Omega_2=(2\pi)\,0.9$ MHz. We record a sinusoidal trace with frequency $\Omega_{2}$, which decays with $T_{2\rho,\text{standard-DD}}\approx\,256\,\mu s$  (see Fig. \ref{coherence_improvement}(a-c)). We shift the modulation frequency of the second field for correlated DD, from $\Omega_1$, experimentally finding the optimal $\widetilde{\Omega}_1=(2\pi)\,4.697$ MHz \cite{Supplemental}. It corresponds to $c\approx 1$ in Eq. \eqref{corr_omega1_MT}, with the $c \rightarrow c+\frac{1}{4}$ correction, as expected from theory for highly correlated noise. 
The coherence time reaches $T_{2\rho,\text{correlated-DD}}\approx\,2.8$ ms -- an improvement of more than an order of magnitude over standard DD (Fig. \ref{coherence_improvement}). It is also $20\%$ higher than the widely used XY8 pulsed dynamical decoupling sequence, which gives 2.32 ms \cite{Supplemental}. Given the proximity to the single-drive coherence time limit (cf. $T_{1\rho}\approx 3$ ms \cite{krantz2019quantum, wang2020coherence}), it is informative to estimate the relaxation-free coherence time of the two schemes, which demonstrates a greater improvement of $67\%$ \cite{Supplemental}. 
We note that this is achieved without optimizing the Rabi amplitudes, which can further prolong the coherence time \cite{ezzell2022dynamical,Supplemental}. In addition, continuous dynamical decoupling has several advantages over pulsed methods, such as uninterrupted protection, negligible memory access latency, and typically lower peak power \cite{khodjasteh2013designing,miao2020universal}.

\begin{figure}[t!]
\centering
\includegraphics[width=0.9\linewidth]{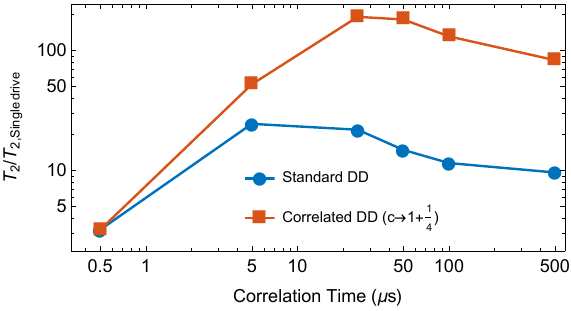}
\centering
\caption{Simulation of the improvement of coherence time $T_{2\rho}$ for standard and correlated DD in comparison to the single-drive $T_{2\rho}\approx 22.3\,\mu$s for different noise correlation times \cite{Supplemental}.
}
\label{Fig:SDD_CDD_comparison_corr_times}
\end{figure}

To further support our findings, we compare standard and correlated DD by a numerical simulation of a qubit subject to environmental noise, typical of NV centers, and correlated field noise, modeled with Ornstein-Uhlenbeck processes \cite{CaiNJP2012,AharonNJP2016,UhlenbeckRMP1945,GillespieAJP1996,GillespieAJP1996a,Supplemental}. Correlated DD achieves a coherence time of $3.8$ ms, an improvement of twenty times over standard DD \cite{Supplemental}. The simulation does not account for relaxation and uses lower Rabi frequencies, resulting in longer coherence times and greater improvement than in the experiment, highlighting the potential capabilities of correlated DD. Figure \ref{Fig:SDD_CDD_comparison_corr_times} demonstrates the improvement of correlated and standard DD compared to single-drive decoupling for different correlation times of the amplitude noise. Correlated DD outperforms the standard scheme for all correlation times (and corresponding noise spectra), highlighting its broad applicability. The improvement is greatest when the correlation time is in the range $(2\pi/\Omega_1,2\pi/\Omega_2)$.

\paragraph{Quantum Sensing.---} 
Standard DD has been used for sensing high-frequency (GHz) \cite{StarkNatComm2017} and low-frequency (sub-MHz) \cite{https://doi.org/10.48550/arxiv.2207.06611} signals with NV centers. 
The sensitivity, typically limited by photon-shot noise, effective phase accumulation rate, and coherence time \cite{taylor2008high,barry2020sensitivity, rondin2014magnetometry, StarkNatComm2017,Degen2017RMP,Supplemental}, reads
\begin{equation}\label{Eq:sensitivity_MT}
    \eta(\tau)= \frac{2}{\gamma_{\text{NV}}\alpha C(\tau) \sqrt{N_{\text{ph}} \tau}},
\end{equation}
where the optimal measurement time $\tau\approx\,T_{2\rho}/2$ \cite{Degen2017RMP,barry2020sensitivity,Supplemental}, $\gamma_{\text{NV}}/2\pi$=28 Hz/nT is the gyromagnetic ratio of the NV electron spin, $C(\tau)$ is the signal contrast at time $\tau$ \cite{Supplemental}, and $N_{\text{ph}}$ represents the average number of photons 
per measurement. The attenuation factor $\alpha$ quantifies the effective phase accumulation rate $g^\prime=\alpha g_0$, where $g_0$ is the amplitude of the sensed field \cite{Degen2017RMP,wang2021nanoscale,Supplemental}. One challenge for DD based sensing is this signal attenuation, that is, having a low $\alpha$. 

Previous work \cite{StarkNatComm2017} has demonstrated sensing of a high-frequency ($\omega_g\sim$ GHz) signal by meeting the resonance condition $\omega_g=\omega_0-\widetilde{\Omega}_1-\Omega_e$, where $\Omega_e\equiv\sqrt{\Omega_2^2+(\Omega_1-\widetilde{\Omega}_1)^2}$, which has an attenuation factor $\alpha=\frac{1}{4}$. We label this the ``high-attenuation'' scheme. We demonstrate a ``low-attenuation'' alternative with $\alpha=\frac{1}{2}$, a twofold improvement in sensitivity, using the resonance condition $\omega_{g}=\omega_{0}-\Omega_e$ \cite{Supplemental}. This improvement also confers an advantage for quantum computing schemes that rely on dressed qubits \cite{CaiNJP2012}.

\begin{figure}[t!]
\centering
\includegraphics[width=1\linewidth]{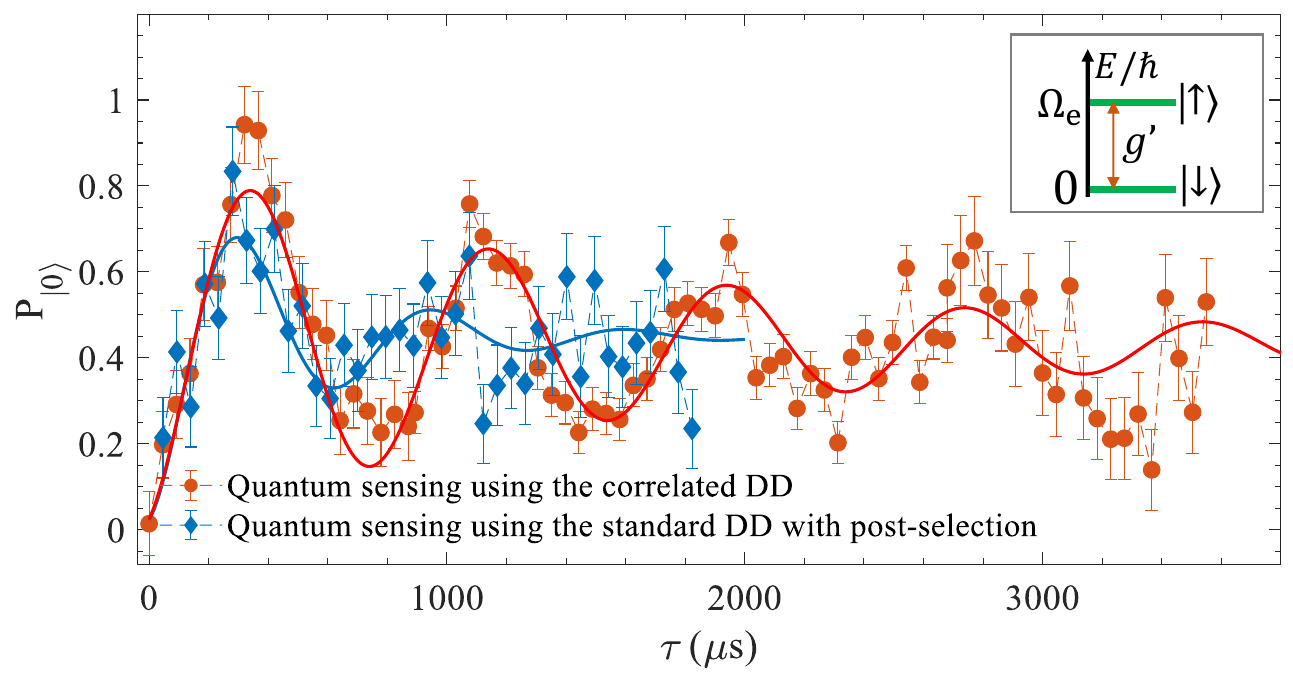}
\centering
\caption{Quantum sensing measurements of an external high-frequency ($(2\pi)$1.487 GHz) signal of amplitude $g_0$ using standard and correlated DD with the ``low-attenuation'' scheme, demonstrating improved sensitivity. We measure the NV state stroboscopically at multiples of $\tau_{\Omega_2}=2\pi/\Omega_{2}$, observing Rabi oscillations with angular frequency $g^\prime\approx g_0/2$ in the lab frame. Standard DD's susceptibility to amplitude noise necessitated data post-selection. \emph{Inset:} a schematic of sensing an external signal using the DD protocols.}
\label{sensing}
\end{figure}

Figure \ref{sensing} shows a comparison of standard and correlated DD for quantum sensing. The parameters are $\Omega_1=(2\pi)$4.350 MHz, $\Omega_2=(2\pi)$0.863 MHz and $\widetilde{\Omega}_1=(2\pi)$4.523 MHz.
Standard DD yields a coherence time of $T_{2\rho}$=494 $\pm 280 ~\mu$s and an effective signal amplitude of $g^\prime=(2\pi)1.541\pm 0.223$ kHz. Correlated DD yields $T_{2\rho}$=1.682 $\pm0.57~$ms and $g^\prime=(2\pi)1.251\pm0.032$ kHz. We note that $\widetilde{\Omega}_1$ does not include the $c\rightarrow(c+\frac{1}{4})$ correction, resulting in slightly lower coherence times than the quantum memory experiments. Standard DD's susceptibility to amplitude noise necessitated data post-selection (about $33\%$ was used) to detect signal-induced oscillations, increasing threefold the total measurement time. This was not necessary for correlated DD due to its robustness. We estimate a correlated DD photon-shot noise limited sensitivity of $\eta\approx 13\; \text{nT/}\sqrt{\text{Hz}}$, which is approximately 3.3 times better than with standard DD due to the longer coherence time and less overhead \cite{Supplemental}. To our knowledge, this sensitivity is better than the state-of-the-art values for high-frequency (GHz) sensing, which are typically in the range of a few hundred \cite{StarkNatComm2017,meinel2021heterodyne,wang2015high} to several tens of $\;\text{nT/}\sqrt{\text{Hz}}$ \cite{staudenmaier2021phase}. Further refinement in terms of Rabi frequency optimization \cite{wang2020coherence} and photon collection efficiency, can improve sensitivity further.

\begin{figure}[t]
\centering
\includegraphics[width=1\linewidth]{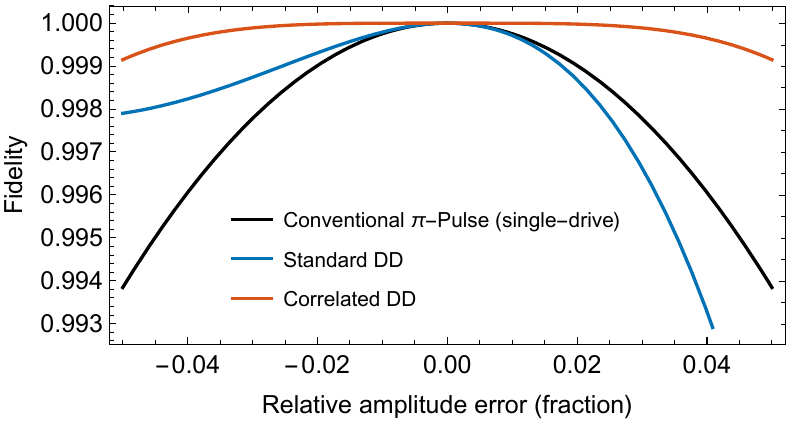}
\centering
\caption{Simulation of fidelity of population transfer for: conventional $\pi$-pulse (using a single-drive, black, pulse time $T=\frac{\pi}{\Omega_1}$), standard DD (using $H_{II}$ in Eq. \eqref{Eq:HII_MT} and $\widetilde{\Omega}_1=\Omega_1$, blue, pulse time $=4T$,$\frac{\Omega_2}{\Omega_1} = 0.25$) and correlated DD (using $H_{II}$ in Eq. \eqref{Eq:HII_MT} and $\widetilde{\Omega}_1=\Omega_1+\frac{\Omega_2^2}{\Omega_1}$, red, pulse time $=3.75T$, $\frac{\Omega_2}{\Omega_1} = \frac{1}{\sqrt{15}}\approx0.258)$. Correlated DD shows superior robustness.}
\label{Pulse_Comparison}
\end{figure}

\paragraph{Robust Coherent Control.---}
Coherent control manipulates quantum systems but noise reduces its fidelity. We demonstrate improved robustness to control noise with correlated DD, compared to a conventional $\pi$-pulse (based on the single-drive) and standard DD, in a population transfer simulation in Fig. \ref{Pulse_Comparison}. Note that the speed of operations for each protocol, i.e. the inverse of the pulse duration, is proportional to the effective Rabi frequency of the dressed qubit. Since the amplitude noise scales with the latter, standard DD prolongs the pulse duration and the coherence time by the same factor, compared to the single-drive pulse. This results in similar robustness for both protocols, as evident in the figure. On the contrary, correlated DD improves the aforementioned scaling by using destructive interference of errors, which results in better robustness.

\paragraph{Discussion ---}\label{Section:Discussion}


Noise cross-correlations emerge naturally in experiments and should be taken into account \cite{almog2016dynamic,Supplemental}. Examples include amplitude noise and its respective Bloch-Siegert shift, noise in the flux-bias and drive amplitude of a flux qubit \cite{yan2013rotating}, and noisy energy shifts and splittings in multilevel systems using multiple control tones \cite{AharonNJP2016,zalivako2023continuous}. Cross-correlations are also present in qubit spatial ensembles addressed by global control fields. Specifically, compensation of amplitude ($B_1$) inhomogeneity in NV ensembles is feasible with correlated DD. Spatiotemporal environmental cross-correlations and cross-talk in coupled multi-qubit systems \cite{von2020two,bradley2019ten} also offer possibilities for optimizing multi-qubit control by destructive interference of correlated noise.

\paragraph{Conclusion ---}\label{Section:Conclusion}

In this Letter, we developed and demonstrated experimentally a destructive interference-based noise protection strategy, which relies on the cross-correlation of two noise sources. We achieve an order-of-magnitude extension of coherence times, allowing for longer quantum memories, improved coherent control, and state-of-the-art sensitivity of quantum sensing. Moreover, correlated DD expands the range of applicable control amplitudes, since such protocols are typically upper-limited by control noise. Implications include higher dynamic range for sensing protocols, faster operations, and improved protection.

Due to its generality, the proposed scheme is applicable to a wide range of physical systems, including trapped atoms and ions, solid-state defects, and superconducting qubits. Our protocol can be combined with and serve as a building block for refocusing-based methods. For example, using it within a rotary echo sequence can correct for the slow spectral components of the uncorrelated part of the amplitude noise, if such exists. 
Composite pulses, pulsed dynamical decoupling, and optimal control can in principle also be optimized when cross-correlated noise is present, utilizing our destructive interference based method. 

\textit{Acknowledgments} --- 
We thank Matthew Markham (Element 6) for the fabrication of the diamond solid immersion lens and Christian Osterkamp (Ulm University) for the  growth of isotopically enriched NV doped layer of CVD diamond. Q.C. is thankful to Raul Gonzalez Cornejo and Gerhard Wolff for experimental help and Yu Liu for helpful discussions. A.S., A.R., and G.G. thank Nati Aharon for fruitful discussions. A.S. gratefully acknowledges the support of the Clore Israel Foundation Scholars Programme, the Israeli Council for Higher Education, and the Milner Foundation. This work was supported by DFG projects 445243414, 499424854 (QUANTERA), 387073854, 386028944 (Forschungsgroßgeräte), 491245864 (Großgeräteinitiative), CRC 1279 and Excellence cluster POLiS, as well as BMBF and the European Union’s Horizon 2020 research and innovation program under grant agreement No 820394 (ASTERIQS), ERC Synergy grant HyperQ (Grant No. 856432), and EU projects C-QuENS, SPINUS, FLORIN, QCIRCLE. A.R. acknowledges the support of ERC grant QRES, project number 770929, Quantera grant MfQDS, ISF and the Schwartzmann university chair. J. M. acknowledges the National Natural Science Foundation of China (Grants No. 12161141011). 
A.S. and Q.C. contributed equally to this work. 


\appendix

\clearpage


\section{Supplemental Material}
\tableofcontents


\setcounter{figure}{0}
\renewcommand{\thefigure}{S.\arabic{figure}}
\setcounter{equation}{0}
\renewcommand{\theequation}{S.\arabic{equation}}




\section{Note 1: Theory}
\section{1a: Standard double drive - assuming uncorrelated noise}\label{sec:theory_detailed}

We consider a two-level system, or qubit, with a Hamiltonian ($\hbar=1$) 
\begin{equation}\label{Eq:lab_frame_H}
    \begin{split}
       H=
       &
       \frac{1}{2}(\omega_0+\delta(t))\sigma_{z} \\
       &
       +\Omega_1(1+\epsilon_1(t)) \cos{(\omega_0 t)}\sigma_{x} \\
       &
       -2\Omega_2 (1+\epsilon_2(t))\sin{\left(\omega_0 t\right)}\cos{(\widetilde{\Omega}_1 t)}\sigma_{x},
    \end{split}
\end{equation}
where $\omega_0$ is the bare energy gap, or transition frequency, of the qubit and $\delta(t)$ is an environment-induced noise term that causes decoherence in the absence of control fields (the second and third summands). 
To protect against noise, we apply a resonant driving field in a perpendicular direction to the noise \cite{CaiNJP2012} (the second summand). Its respective Rabi frequency is $\Omega_1$ with $\epsilon_1(t)$ -- a relative error due to field noise. The motivation for using a relative amplitude error to model drive noise is due to its experimental nature, that is, it is typically proportional to the driving field strength (\cite{kong2018nanoscale} and Fig. \ref{rabi_decay}).

In order to compensate the noise in the first driving field, we apply a second field, the third summand in Eq. \eqref{Eq:lab_frame_H}, which is characterized by a modulation frequency $\widetilde{\Omega}_1$, a Rabi frequency $\Omega_2$, and a relative amplitude error $\epsilon_2(t)$ due to its noise.  The relative, zero-mean, noise terms are assumed to be stationary and have equal and sub-unity variance ($\overline{\epsilon_i(t)}=0, \sigma^2\equiv\overline{\epsilon_i(t)^2}\ll1$, overbar indicates expected value or average over noise realizations or experimental runs). For the standard double drive (DD) protocol (which does not account for potential cross-correlations in the noise of the control fields, in practice assuming $\overline{\epsilon_1(t)\epsilon_2(t)}=0$), the modulation frequency of the second field is set on resonance with the first Rabi frequency, namely, $\widetilde{\Omega}_1=\Omega_1$\cite{CaiNJP2012,StarkNatComm2017}.

In order to understand the noise suppression mechanism of the standard DD protocol, we first consider the stability of the bare energy gap without control fields
\begin{equation}\label{Eq:Gap_E0}
    \Delta E_0(t)=\omega_0+\delta(t).    
\end{equation}

In many experiments the system is prepared in an initial maximum superposition state, e.g. $|\psi(t=0)\rangle=\frac{1}{\sqrt{2}}(|0\rangle+|1\rangle)$, and the state evolves in time to $|\psi(t)\rangle=\frac{1}{\sqrt{2}}(e^{-i\frac{1}{2}\phi(t)}|0\rangle+e^{i\frac{1}{2}\phi(t)}|1\rangle)$, accumulating a phase difference $\phi(t)=\int_0^t \Delta E_0(t^\prime)\,d t^\prime=\omega_0 t+\int_0^t \delta (t^\prime)\,d t^\prime$. The noise term $\delta (t)$ leads to a variation of the energy gap $\Delta E_0(t)$, and thus to decoherence. The fidelity of the final state is then given by $\overline{\langle \psi_0(t)|\psi(t)\rangle}=\overline{\cos{\int_0^t \delta (t^\prime)\,d t^\prime}}=e^{-\overline{\delta^2} t^2/2}=e^{-\left(t/T_2^\ast\right)^2}$, where $|\psi_0(t)\rangle$ is the wavefunction in the noiseless case ($\delta(t)=0$). In the last two equalities we assumed for simplicity that $\delta(t)=\delta$ is a Gaussian random variable, constant during a single experimental run, with variance $\overline{\delta^2}$, and $T_2^\ast$ is the corresponding coherence time \cite{Suter2016RevModPhys}. It is evident that reducing the variation of the energy gap $\Delta E_0(t)$ would prolong the coherence time of our system, leading to improved quantum memories and better sensitivity for quantum sensing. Therefore, in the following analysis we shall focus on optimizing the stability of the energy gap with respect to the noise, namely, minimizing its variation. In a following section we discuss the conditions under which this criterion analytically leads to prolonging of coherence times, possible generalizations, and when its validity and effectiveness has to be established in numerical trials.

We now examine the noise suppression mechanism of the standard double drive protocol. First, we move to the interaction picture with respect to $H_0^{(1)}=\omega_0 \sigma_{z}/2$, and apply the rotating-wave approximation ($\Omega_1,\Omega_2,\widetilde{\Omega}_1\ll\omega_0$). The evolution of the system is described by the Hamiltonian $H_{I}$ in the first interaction picture - 
\begin{equation}\label{Eq:H_I}
    \begin{split}
    H_{I}&=U_0^{(1)\dagger}H U_0^{(1)}-i U_0^{(1)\dagger}\left(\partial_t U_0^{(1)}\right)\approx\frac{\delta(t)}{2}\sigma_{z}+\\
    &\frac{1}{2}\Omega_1(1+\epsilon_1(t))\sigma_{x}+\Omega_2 (1+\epsilon_2(t))\cos{(\widetilde{\Omega}_1t)}\sigma_{y},
    \end{split}
\end{equation}
where $U_0^{(1)}=\exp{(-i H_0^{(1)} t)}$. In the ideal case of a perfect first driving field ($\epsilon_1(t)=0$), the second field is unnecessary ($\Omega_2=0$) and the strong Rabi frequency $\Omega_1$ suppresses the effect of the noise $\delta(t)$ \cite{CaiNJP2012}.  This is evident by examining the energy gap in the interaction picture in the absence of a second drive ($\Omega_2=0$), which is given by 
\begin{equation}\label{Eq:CDD_1_egap}
    \Delta\,E_I(t)=\sqrt{\Omega_1^2+\delta(t)^2}\approx \Omega_1+\frac{\delta(t)}{2\Omega_1}\delta(t)
\end{equation}
where the last approximation is valid when $|\delta(t)|\ll\Omega_1$. It is evident that the noise-induced detuning $\delta(t)$ in the bare basis is reduced by a factor $\frac{\delta(t)}{2\Omega_1}$ in the interaction basis due to the effect of the strong first driving field, leading to a longer coherence time. 
However, in real experimental situations the driving field noise $\epsilon_1(t)$ is non-zero and it itself causes decoherence in the interaction basis as $\Omega_1\rightarrow \Omega_1(1+\epsilon_1(t))$ in Eq. \eqref{Eq:CDD_1_egap}.

We use the second driving field to reduce the effect of the noise of the first field \cite{CaiNJP2012}. In order to understand the mechanism of noise reduction, we transform the Hamiltonian in Eq. \eqref{Eq:H_I} to a second interaction picture with respect to $H_0^{(2)}=\widetilde{\Omega}_1 \sigma_{x}/2$, and apply the rotating-wave approximation once more ($\Omega_2\ll\widetilde{\Omega}_1$). We further assume that the effect of the environment noise $\delta(t)$ can be neglected due to the first strong driving field (see Eq. \eqref{Eq:CDD_1_egap}). Thus, the Hamiltonian in the second interaction picture is
\begin{equation}\label{Eq:HII}
    H_{II}=\frac{\left(\Omega_1-\widetilde{\Omega}_1\right)+\Omega_1\epsilon_1(t)}{2}\sigma_{x}+\frac{\Omega_2}{2}(1+\epsilon_2(t))\sigma_{y}.
\end{equation}
In the case of the standard double drive (which implicitly assumes uncorrelated noise) the modulation frequency of the second driving field $\widetilde{\Omega}_1$ is resonant with the noiseless energy gap due to the first driving field $\Delta E_{I}(t)=\Omega_1$, i.e., $\Omega_1-\widetilde{\Omega}_1=0$ \cite{CaiNJP2012,StarkNatComm2017}. Then, the energy gap in the second interaction picture is given by
\begin{equation}\label{Eq:CDD_2_egap}
\begin{aligned}
    \Delta E_{II}(t)&=\sqrt{\Omega _2^2 \left(1+\epsilon _2(t)\right){}^2+\Omega _1^2 \epsilon_1(t)^2}\\
    &\approx 
    \Omega_2 \left(1+\epsilon _2(t)\right)+\frac{\Omega _1 \epsilon _1(t)}{2 \Omega _2}\Omega _1 \epsilon _1(t)
\end{aligned}
\end{equation}
where the last approximation is valid when $|\Omega_1\epsilon_1(t)|\ll\Omega_2$. For a perfect second drive ($\epsilon_2(t)=0$), it is evident that the amplitude error of the first drive $\Omega _1 \epsilon _1(t)$ is reduced by a factor $\sim\frac{\Omega _1 \epsilon _1(t)}{2 \Omega _2}$. 
However, in real experimental situations the second driving field noise $\epsilon_2(t)$ is non-zero and it itself causes decoherence in the second interaction picture due to the error term $\Omega_2\epsilon_2(t)$ in Eq. \eqref{Eq:CDD_2_egap}, even if the effect of $\epsilon_1(t)$ becomes negligible. In this case, the error rate is much smaller than with the first drive only as $\Omega_2\epsilon_2(t)\ll\Omega_1\epsilon_1(t)$ because $\Omega_2\ll\Omega_1$ but it still leads to uncompensated decoherence.

Importantly, while the absolute error rate is smaller, the relative errors remain similar since $\epsilon_1(t)$ and $\epsilon_2(t)$ are usually of the same magnitude. Namely, the coherence time is prolonged and the effective energy gap is decreased by the same factor. Thus, the speed of the basic operation has to be reduced in order to prolong coherence. 
As the relative error remains unchanged, $\pi$-pulse fidelities, for example, do not improve by using the double-drive scheme, as we show in a subsequent note.
Generally, the importance of absolute error rates vs. relative ones depend on the task at hand (e.g. quantum memory, sensing, gates etc).

\section{1b: Correlated double drive - accounting for potential noise cross-correlations }\label{sec:theory_detailed_corr}

Next, we modify the double drive protocol to account for potential cross-correlations in the amplitude noise of the two driving fields. Importantly, we show that when such correlations exist (which is the case in our experimental setup) both the absolute and relative error rates improve. While both improvements result from better noise protection, the latter further benefits from a larger effective energy gap. Furthermore, the magnitude of the second Rabi frequency $\Omega_2$ in the standard DD protocol is limited due to two effects. First, the rotating wave approximation requires $\Omega_2\ll\widetilde{\Omega}_1(=\Omega_1)$\cite{Yudilevich_2023}, and second, as $\Omega_2$ increases so does the associated noise term $\Omega_2\epsilon_2(t)$. The new protocol mitigates both effects, thereby allowing the use of higher values of $\Omega_2$. This in turn allows for a wider range of $\Omega_1$, expanding the parameter regime in which the scheme has long coherence time as well as it's range for effective spectroscopy and coupling to other quantum systems. Moreover, higher values of $\Omega_2$ allow for better dynamical decoupling of the $\epsilon_1(t)$ noise, as shown previously.

As explained in the previous section, the double-drive protocol is a natural generalization of the single-drive protocol. The choice of resonant drive in the first interaction picture ($\widetilde{\Omega}_1=\Omega_1$), follows naturally from the single-drive logic. We show, however, that this choice is not optimal when the relative error terms, $\epsilon_1(t)$ and $\epsilon_2(t)$ are correlated. Explicitly, we define the 
cross-correlation of the control noise terms
\begin{equation}
    c\equiv\frac{\overline{\epsilon_1(t)\epsilon_2(t)}}{\sigma^2},
\end{equation}
and optimize the drive frequency for any value $c$ may take.
Indeed, when noise terms result from different underlying physical processes and systems, as is the case for $\delta(t)$ and $\epsilon_1(t)$ in our experimental setup, they are typically not correlated \cite{almog2016dynamic}. The situation for $\epsilon_1(t)$ and $\epsilon_2(t)$ is different since, despite the fact that they correspond to different spectral components of the drive, both terms share control hardware. We show that if such noisy fluctuations are correlated, it is possible to \textit{destructively interfere} them on the qubit by detuning the second drive's modulation frequency ($\widetilde{\Omega}_1$) from resonance, thereby mitigating their effect and significantly prolonging the qubit's coherence time. 

The Hamiltonian in Eq. \eqref{Eq:HII} gives rise to the (doubly) dressed-state picture of two-level systems \cite{cohen1994atoms}. It can be made more explicit by rotating our basis by an angle $\nu=\frac{1}{2}\arctan(\frac{\Omega_2}{\Omega_1-\widetilde{\Omega}_1})$, with the transformation $U_0^{(3)}=\exp{(-i \nu\sigma_z)}$, 
\begin{equation}
    H_{IIS} = -\frac{\Omega_e}{2}\sigma_x,
\end{equation}
where  $\Omega_e\equiv\sqrt{\Omega_2^2+(\Omega_1-\widetilde{\Omega}_1)^2}$, and the noise terms were omitted for simplicity of presentation. Similarly to bare qubits, dressed qubits can act as fundamental building blocks for quantum information processing in sensing and computing applications \cite{CaiNJP2012}. 

As we explain in detail in the following sections, our destructive-interference based protocol creates a dressed qubit (a combined system of the bare qubit and the drive) with superior stability that is not simply inherited from the individual stabilities of its separate components.

\paragraph{Perfect correlation.---}

The main novelty of our scheme can be demonstrated by considering the simple, yet most advantageous, case of perfect correlation -  $c=1$. Under the zero-mean and equal variance assumptions, perfect correlation implies $\epsilon_1(t)=\epsilon_2(t) \equiv\epsilon(t)$. Note that the total control Hamiltonian (in the lab-frame) has the following form - $\sim f(t)*(1+\epsilon(t))\sigma_x$, namely, a relative noise model for the entire waveform independent of its spectral composition. The energy gap of the second interaction picture Hamiltonian (Eq. \eqref{Eq:HII}) reads

\begin{equation}\label{Eq:Corr_gap}
\small
\begin{aligned}
    \Delta \widetilde{E}_{II}(t)&=\sqrt{\Omega _2^2 \left(1+\epsilon _2(t)\right){}^2+((\Omega_1-\widetilde{\Omega}_1)+\Omega_1\epsilon_1(t))^2}\\
    &\approx 
    \Omega_e + \frac{\Omega _2^2\epsilon_2(t)+\Omega
   _1(\Omega
   _1-\widetilde{\Omega}_1)\epsilon_1(t)}{\Omega_e},\\
   &=
   \Omega_e + \frac{\Omega _2^2+\Omega
   _1(\Omega
   _1-\widetilde{\Omega}_1)}{\Omega_e}\epsilon(t),
\end{aligned}
\end{equation}
where each summand in the numerator of the prefactor of $\epsilon(t)$ corresponds to noise from a different driving field (which were set equal).

We are now at a position to observe that by shifting the second frequency according to
\begin{equation}\label{Eq:Shift_Corr_1}
    \widetilde\Omega_1=\Omega_1+\frac{\Omega_2^2}{\Omega_1},
\end{equation} 
the two summands have equal magnitude and opposite sign. Thus, noise from the two driving fields destructively interferes to nullify the term completely, thereby stabilizing the energy gap to first order. Note that this correlation-induced frequency shift is unrelated to the Bloch-Siegert shift \cite{bloch1940magnetic,james2007effective}. It is of different magnitude and physical origin, namely, the former exists in the case of circularly polarized control fields as well. A combined treatment of both effects appears in a following section.

Adding the next term to the expansion of the energy gap in Eq. \eqref{Eq:Corr_gap}, it now reads
\begin{equation}\label{CNSGap}
\begin{aligned}
    \Delta \widetilde{E}_{II}(t)&\approx 
    \Omega_e + 
    \frac{\Omega_1 \sqrt{\Omega_1^2+\Omega_2^2}}{2\Omega_2}\epsilon(t)^2 ,
\end{aligned}
\end{equation}
and the noise term appears only to second order. Thus, in the case of perfect correlation, the protocol makes the doubly-dressed states (dynamic) clock states. This can be seen geometrically as well, by observing the $\widetilde{\Omega}_1$ dependent angle between the Bloch vectors of the effective drive and the effective noise. To do so, we transform $H_{II}$ once more using (the $\widetilde{\Omega}_1$ dependent) $U_0^{(3)}$, which results in 
\begin{equation}\label{H_IIS}
\begin{aligned}
    H_{IIS} = 
    -\frac{\Omega_e}{2}
   \sigma_x
   +
   \sqrt{\Omega_1^2+\Omega_2^2}\epsilon(t)\sigma_y.
\end{aligned}
\end{equation}
Thus, noise suppression is facilitated by orienting the effective drive in a perpendicular direction to the effective correlated noise, thereby nullifying the projections of the noise (which have the first order effect on decoherence) on the effective drive axis. This is in clear contrast to the standard DD case where $\widetilde{\Omega}_1=\Omega_1$ and the effective noise in the Hamiltonian in Eq. \eqref{Eq:HII} has both parallel and perpendicular components to the effective drive.

\paragraph{Partial correlation.---}

While the 
cross-correlation satisfies $|c|\leq1$, it not need equal $1$, of course. Going back to the expression in the second line of Eq. \eqref{Eq:Corr_gap}, we can proceed in an analogous way to the previous section by minimizing its standard deviation. Direct calculation yields
\begin{equation}\label{Eq:STD}
\begin{aligned}
    &\sqrt{\text{Var}(\Delta \widetilde{E}_{II}(t))}\approx
    \sqrt{\overline{\left(\frac{\Omega _2^2\epsilon_2(t)+\Omega_1(\Omega_1- \widetilde{\Omega}_1)\epsilon_1(t)}{\Omega_e}\right)^2}}
    \\& 
    =\sigma\sqrt{\left(
    \frac{\Omega_1^2(\Omega_1-\widetilde{\Omega}_1)^2+\Omega_2^4}{(\Omega_1-\widetilde{\Omega}_1)^2+\Omega_2^2}
    +
    c \frac{2 \Omega_1 \Omega_2^2(\Omega_1-\widetilde{\Omega}_1)}{(\Omega_1-\widetilde{\Omega}_1)^2+\Omega_2^2}
    \right)},
\end{aligned}
\end{equation}
which can then be minimized with respect to $\widetilde{\Omega}_1$. 
The optimal frequency reads 
\begin{equation}\label{corr_omega1}
\small
    \begin{aligned}
    \widetilde{\Omega}_1
    &
    =\frac{\sqrt{4c^2\Omega_2^2\Omega_1^2+
    (\Omega_2^2-\Omega_1^2)^2}+2c\Omega_1^2+(\Omega_2^2-\Omega_1^2)}{2c \Omega_1}
    \\&
    \approx
    \Omega _1+c \frac{\Omega_2^2}{\Omega_1},\\
    \end{aligned}
\end{equation}
where in the second line we have assumed $\Omega_2\ll\Omega_1$. Note that the result reproduces the perfect correlation ($c=1$) case even without the approximation, as well as recovering the zero correlation case ($c=0$) of the standard DD easily.
The minimal standard deviation (Eq. \eqref{Eq:STD} subject to Eq. \eqref{corr_omega1}) reads, to leading order in $\Omega_2\ll\Omega_1$,
\begin{equation}\label{Eq:minSTD}
    \sqrt{\text{Var}(\Delta \widetilde{E}_{II}(t))}\approx
    \sqrt{(1-c^2)}\sigma\Omega_2,
\end{equation}
and the results of the standard DD protocol and the perfect correlation case are once again clearly recovered. 

Importantly, the clear recovery of the standard DD (for $c=0$) isolates the contribution of the two-field correlation from the standard dynamical decoupling properties of the protocol. Furthermore, it shows that the destructive interference principle is compatible with dynamical decoupling. Thus, we demonstrate and compare the standard and correlated DD schemes experimentally and in simulation.

\paragraph{Correlated-noise and the Bloch-Siegert shift.---}

The resulting form of the frequency shift demands a revisit of the second rotating-wave approximation leading to Eq. \eqref{Eq:HII} ($\Omega_2\ll\widetilde{\Omega}_1$). The next order of this approximation is the Bloch-Siegert shift \cite{bloch1940magnetic}, which is additive in the regime $\Omega_2\ll\widetilde{\Omega}_1$ \cite{james2007effective}. With this final consideration in mind, the final form of optimal frequency reads
\begin{equation}\label{corr_omega_BS}
    \widetilde{\Omega}_1\approx\Omega_1+\left(c+\frac{1}{4}\right)\frac{\Omega_2^2}{\Omega_1}.
\end{equation}

The above expression for the \textit{correlated-noise-shift} is the main result of this section.

\section{1c: Coherence optimization via energy gap stabilization}

In the previous sections, we have minimized the variation of the energy gap (i.e., stabilized it) to optimize the coherence time. The reasoning was given in the example following Eq. \eqref{Eq:Gap_E0}. Let us generalize this example to an initial-state independent case, and explicitly specify the assumptions that lead to the gap stabilization criterion.

We define an initial-state $(|\psi\rangle)$ independent fidelity measure following \cite{BowdreyPhLett2002} (using a common fidelity measure for the synthesis of unitary gates \cite{wilhelm2020introduction} results in a similar final expression) as
\begin{equation}\label{eq:Fidelity}
\small
    F(t)=\overline{\frac{1}{4\pi}\int 
    \text{Tr}(\widetilde{U}(t,t_0) |\psi \rangle\langle\psi |\widetilde{U}(t,t_0)^{\dagger}\mathcal{M}[|\psi \rangle\langle\psi|])
    d S},
\end{equation}
where $\widetilde{U}(t,t_0)$ is the time-evolution operator, potentially in one of the interaction pictures, the integration is over the surface $S$ of the Bloch sphere, overbar indicates average over different noise-realizations, or experiments, and for quantum memory we set $\mathcal{M}=\mathbb{1}$. The initial-state independent coherence time ($T_2^{\text{ISI}}$) is defined as the time it takes the envelope of $F(t)$ to decay from $1$ to $\approx 0.79$, i.e., to a $1/e$ drop in the difference to $0.67$, which is the decay limit when we neglect population relaxation (in the dressed basis).

We note that unlike coherence times that are defined by CPMG ($T_2^{\text{CPMG}}$) or spin-lock ($T_{1\rho}$) experiments, the initial-state-independent approach is more suitable for quantum computing and memory applications, as well as for some sensing applications \cite{Degen2017RMP}.

The first assumption we make is that the Hamiltonian commutes with itself at different times during a single experimental run. Except for the case discussed previously (following Eq. \eqref{Eq:Gap_E0}), we can also consider the case of Eq. \eqref{Eq:HII} where amplitude fluctuations are assumed to be slow compared to the time of a single experiment. In this case, the Hamiltonian can be approximated as constant over a single experimental run (yet noisy between experiments).

Under this assumption, the fidelity (Eq. \eqref{eq:Fidelity}) can be greatly simplified -  
\begin{equation}\label{SimlifiedFidelity}
    F(t)=\frac{1}{3}(2 + \overline{\cos(\phi(t))}),
\end{equation}
where $\phi(t)=\int_0^t\Delta \widetilde{E}(t') dt'$ is the accumulated phase, and $\Delta\widetilde{E}(t)$ is the Hamiltonian's energy gap. For the envelope of $F$ to decay $\phi(t)$ has to vary, such that oscillations of different noise realizations average out. Thus, it is evident that the smaller the variance of the phase - the longer the coherence time. This is made analytically precise when $\phi(t)$ has Gaussian statistics, which we shall count as a second assumption, though not a stringent one. For example, this is the case when the noise $\epsilon_i(t)$ is Gaussian and the energy gap is expanded to the first order in Eq. \eqref{Eq:Corr_gap}. In this case, the average can be calculated analytically and the variance $\overline{\phi(t)^2}=\int_0^t\int_0^t\Delta \widetilde{E}(t')\Delta \widetilde{E}(t'')dt'dt''$ determines the decay. This expression contains the autocorrelation of the noise as well as cross-correlation between different noise terms. Zero and finite lag ($\equiv t'-t''$) contributions exist for both types of correlations. This can be seen, for example, by substituting the second line of Eq. \eqref{Eq:Corr_gap} into the expression. The auto-correlation terms, and the power spectral densities which are their Fourier transforms,  are a key component in the analysis of dynamical decoupling. It is clear, therefore, that the cross spectral densities, namely, the Fourier transform of the cross-correlations, can similarly play a significant role in the full analysis.

The third assumption that is needed to single out the contribution of the zero-lag cross-correlation and auto-correlation terms is the time independence of the energy gap. In this case $\overline{\phi(t)^2}=\overline{\Delta\widetilde E ^2}t^2$ and only the aforementioned terms survive. This formalizes the intuitive reasoning presented previously, namely that stabilization of the energy gap prolongs the coherence time.

When these assumptions do not strictly hold the validity and effectiveness of the stability criterion has to be established in numerical trials. As long as decoherence is dominated by the stability of the energy gap, we expect our approach to deliver significant advantages. In this work we show by numerical simulations that this is indeed the case, although none of the aforementioned assumptions strictly hold and experimentally realistic noise models are used. The simulations include environmental noise as well, and the results verify that the protocol maintains its dynamical decoupling properties. 

\section{Note 2: Sample and Setup}\label{Section:Setup}

We perform our experiments in a single nitrogen-vacancy (NV) center. 
The NV center is a point defect in the diamond lattice, consisting of a substitutional nitrogen atom and a vacancy on the neighboring lattice site. It has a $^3\mathrm{A}_2$ triplet ground state with a zero-field splitting of $D\approx2.87~\mathrm{GHz}$. The diamond sample was produced by chemical vapor deposition and polished into a hemispherical shape of 2 mm diameter. In order to create NV centres in this diamond, the flat surface was overgrown with an about 100 nm thick layer of isotopically enriched ${}^{12}\!$C (99.999\%) using the plasma enhanced chemical vapor deposition method \cite{osterkamp2015stabilizing}. The diamond's hemispherical shape acts as a solid immersion lens (SIL), enhancing the photon collection efficiency \cite{hadden2010strongly,siyushev2010monolithic}. As a result, the saturated photon flux of a single NV center can reach up to $10^3$ kHz.

Using a home-built confocal setup, the NV spin states can be initialized and read out by a 532 nm laser (Laser Quantum gem 532). A magnetic bias field of 49 mT aligned with the NV axis lifts the degeneracy of the $m_{s}= \pm 1$ spin states, and also effectively polarizes the $^{15}$ N nuclear spins \cite{jacques2009dynamic}, in which case the hyperfine coupling effects of $^{15}$N nuclear spin was removed from the measurements outcome. All control fields were generated with an arbitrary waveform generator (Tektronix AWG70001A, sampling rate 50 GSamples/s) with arbitrary phases and amplitudes. In sensing experiments, the sensed signals were generated with either the AWG for the measurements shown in Fig. \ref{sensing_scheme_results}, or an independent signal generators (Rohde$\&$Schwarz, SMIQ03) (see main text Fig. 4
).
\section{Note 3: Quantum Memory - Additional Experimental Results}
\subsection{3a: Ramsey and Rabi decay}\label{Section:rabi}

\begin{figure}[t]
\centering
\includegraphics[width=1\linewidth]{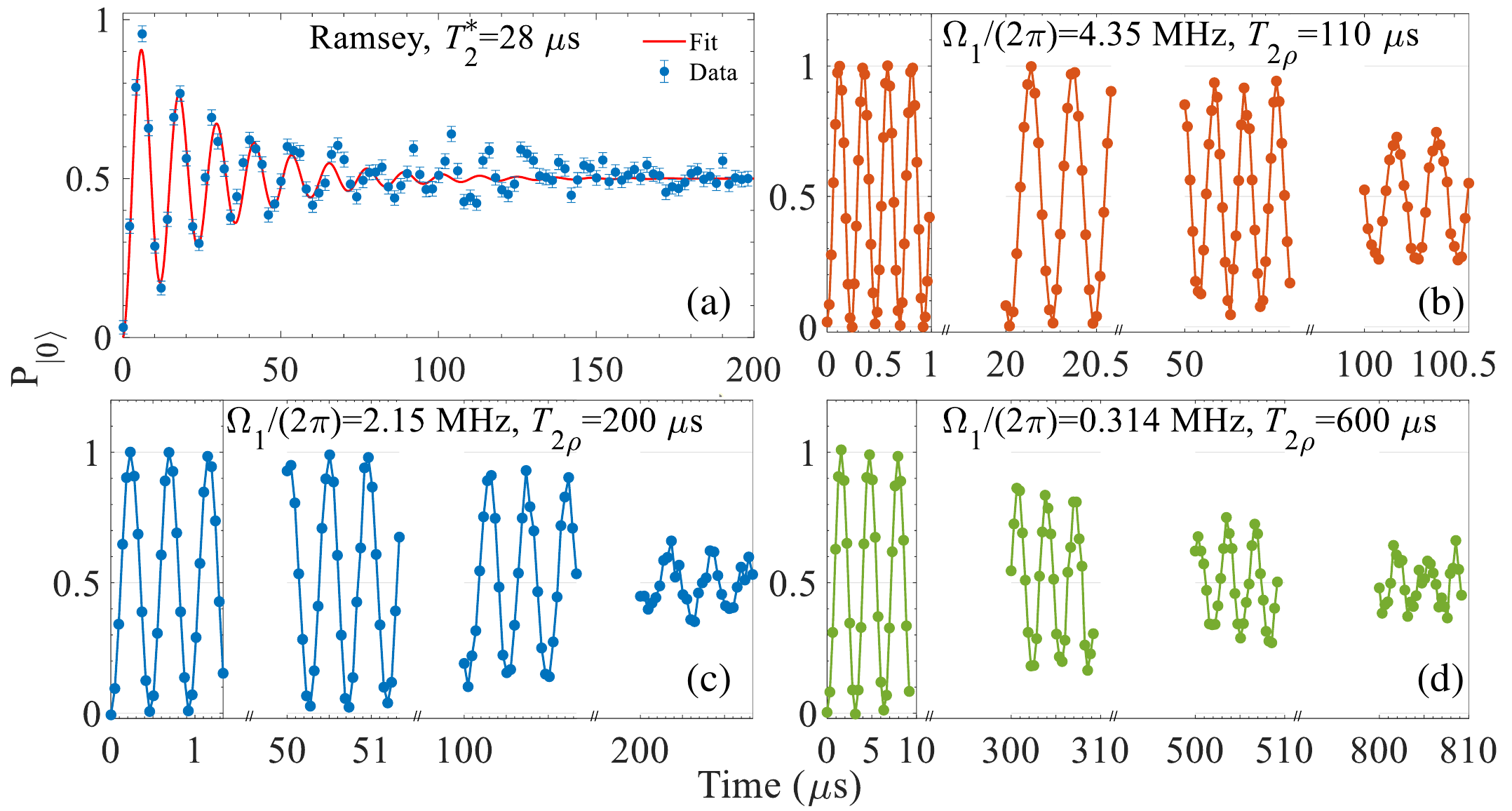}
\centering
\caption{Experimental measurements of the population of state $|0\rangle$ ($P_{|0\rangle}$) vs. experimental time for (a) a Ramsey measurement, yielding a pure dephasing time of $T_{2}^{\ast}=28\;\mu$s and Rabi oscillations measurements with three different Rabi frequencies: $\Omega_1/(2\pi)$ equal to 4.35 MHz, 2.15 MHz, and 0.314 MHz for (b), (c), and (d), respectively. The resulting Rabi decay times were 110 $\mu$s, 200 $\mu$s, and 600 $\mu$s, respectively.}
	\label{rabi_decay}
\end{figure}

We perform Ramsey and Rabi decay measurements to calibrate the expected noise. Fig. (\ref{rabi_decay}.a) shows the experimental results from a Ramsey measurement with pure dephasing time of $T_{2}^{\ast}=28\;\mu$s. We then conduct Rabi measurements at three different Rabi frequencies ($\Omega_1$ =$(2\pi)$4.35 MHz, $(2\pi)$2.15 MHz, and $(2\pi)$0.314 MHz) to investigate the coherence times under a single protecting driving field, as shown in Fig. (\ref{rabi_decay}.b,c,d). The resulting coherence times are 110 $\mu$s, 200 $\mu$s, and 600 $\mu$s, respectively, with higher Rabi frequencies leading to shorter coherence times, indicating an increase in amplitude noise associated with higher Rabi frequencies. The coherence time indeed scales inversely with the Rabi frequency, in the two measurements involving the higher frequencies, as expected in the regime when the amplitude noise is dominant and typically relative (a percent of the Rabi frequency). This scaling is not expected to hold for sufficiently low Rabi frequencies, as magnetic noise decoupling becomes inefficient. Indeed, the low Rabi measurement (with $\Omega_1/(2\pi)$ = 0.314 MHz) doesn't follow the same scaling.

\subsection{3b: Experimental optimization of $\widetilde{\Omega}_1$}

\begin{figure}[t]
\centering
\includegraphics[width=1\linewidth]{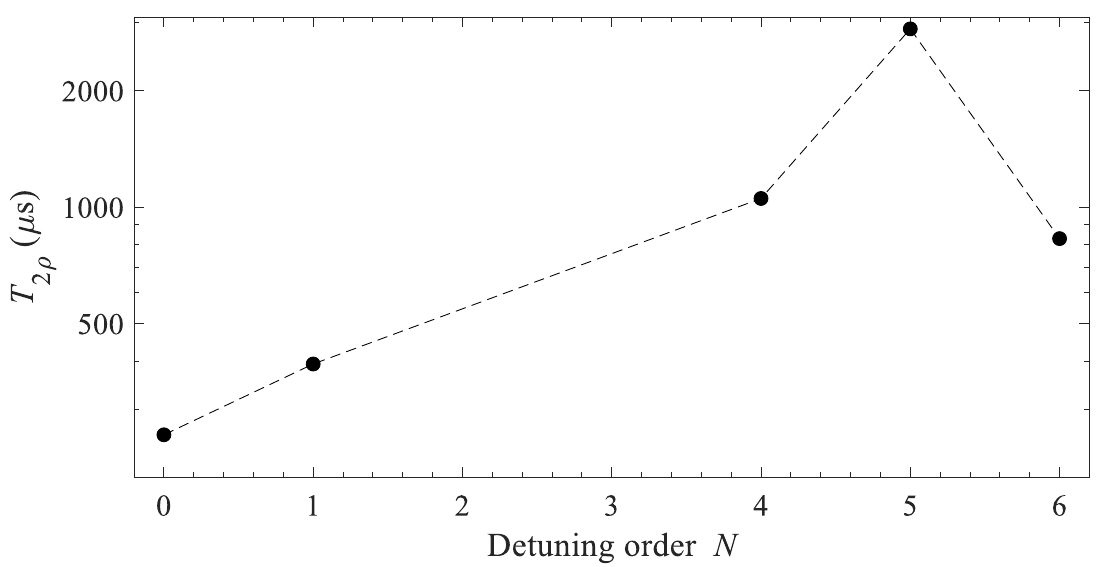}
\centering
\caption{$T_{2\rho}$ obtained using different frequency shifts $\widetilde{\Omega}_1(N)=\Omega_1+\frac{N}{4}\frac{\Omega_2^2}{\Omega_1}$, labeled by $N$. The maximum at $N=5$ suggests a high-degree of amplitude noise cross-correlation ($c\approx\,1$). The experimental values for $\Omega_1$ and $\Omega_2$ are the same as the one in Fig. 2 
in the main text. 
}
	\label{T2rho_with_N}
\end{figure}

In order to find the optimal modulation frequency $\widetilde{\Omega}_1$ of the second driving field, we perform a series of experiments where we measure the coherence times $T_{2\rho}$ of an NV center in diamond for different frequency shifts (i.e. detunings) labeled by $N$, and denoted as 
\begin{equation}\label{Eq:DetuningOrderN}
\widetilde{\Omega}_1(N)=
\Omega_1+\frac{N}{4}\frac{\Omega_2^2}{\Omega_1},
\end{equation} 
from which we can extract the cross-correlation parameter, $c$, according to Eq. \eqref{corr_omega_BS}. The Rabi frequencies applied in the DD experiments are $\Omega_1=(2\pi)$ 4.460 MHz, $\Omega_2=(2\pi)$0.900 MHz. The results are shown in Fig. \ref{T2rho_with_N}. The coherence time is maximal for $N=5$, which is expected by the theory for a high degree of amplitude noise correlation $c \approx 1$, when taking into account the Bloch-Siegert shift (cf. Eq. \eqref{corr_omega_BS}). We also note that in other experiments (not shown) the improvement of the coherence time for $N=5$ in comparison to $N=4$ is less pronounced but the overall dependence on $N$ is consistently similar.


\subsection{3c: XY8 measurements}\label{Section:XY8}
To investigate coherence times under pulsed dynamical decoupling sequences, we conducted XY8-$N$($N$ denotes the repetitions of XY8  blocks) measurements \cite{Degen2017RMP}. The Rabi frequency of the applied pulses is $(2\pi)23\;$MHz. 

\begin{figure}[t!]
\centering
\includegraphics[width=1\linewidth]{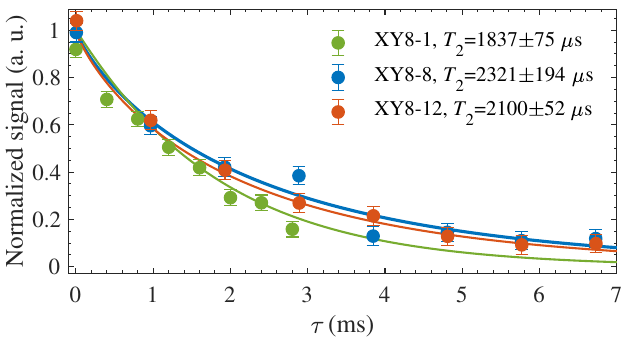}
\centering
\caption{Experimental measurements of the normalized signal difference from two alternating measurements (see text) vs. experimental time $\tau$ for pulsed dynamical decoupling with XY8-$N$ sequences, where $N$ is the number of times the sequence is repeated. $N$ was set to 1,8, and 12 we varied the time between the centers of the pulses $\tau_{p}$, so the experimental time is $\tau=8N\tau_{p}$. We obtain $T_{2}$ values of 1.84 ms, 2.32 ms, and 2.1 ms, respectively. These results suggest that that the coherence time saturates at approximately 2.32 ms.}
	\label{xy8_data}
\end{figure}

In a first set of experiments, we varied the inter-pulse delay, i.e., the time between the centers of the $\pi$ pulses, for each XY8-$N$ sequence. Figure (\ref{xy8_data}) shows the experimental results and indicates that the coherence time is saturated at $T_{2}\approx2.34\;$ms , which is lower than $T_{2\rho}$ obtained by our correlated DD scheme.

\begin{figure}[b]
\centering
\includegraphics[width=1\linewidth]{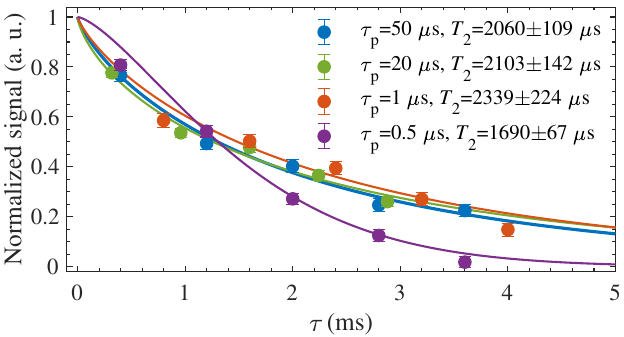}
\centering
\caption{Experimental measurements of the normalized signal difference, obtained from two alternating measurements (see text) vs. experimental time $\tau$ for pulsed dynamical decoupling with XY8-$N$. In contrast to Fig. \ref{xy8_data}, we fix the inter-pulse spacing $\tau_{p}$ for each measurement and increase number of repetitions $N$ of the sequence to change the measurement time $\tau=8N\tau_{p}$. By applying a fit function $\exp(-(\tau/T_{2})^\beta)$ to the data, we obtain the corresponding $T_{2}$ values. }
\label{xy8_N_scan}
\end{figure}

Experiments where we vary the inter-pulse delay $\tau_{p}$ and the number of pulses $N$ is kept constant are typically used to probe the spin environment noise spectrum \cite{Degen2017RMP}. 
However, for quantum memory experiments it is usually preferable to choose an optimal interpulse delay and vary the number of pulses $N$ \cite{ezzell2022dynamical}. To explore the limit of the quantum memory time, we thus vary the order $N$ of the XY8 sequence, keeping $\tau_{p}$ fixed. Figure \ref{xy8_N_scan} shows the respective coherence times for different interpulse delays with the highest $T_2\approx 2.23$ ms, which is similar to the result with correlated DD. 

\subsection{3d: Lab and rotating-frame relaxation - $T_1$ and $T_{1\rho}$ measurements}\label{Section:T1}

\begin{figure}[t]
\centering
\includegraphics[width=1\linewidth]{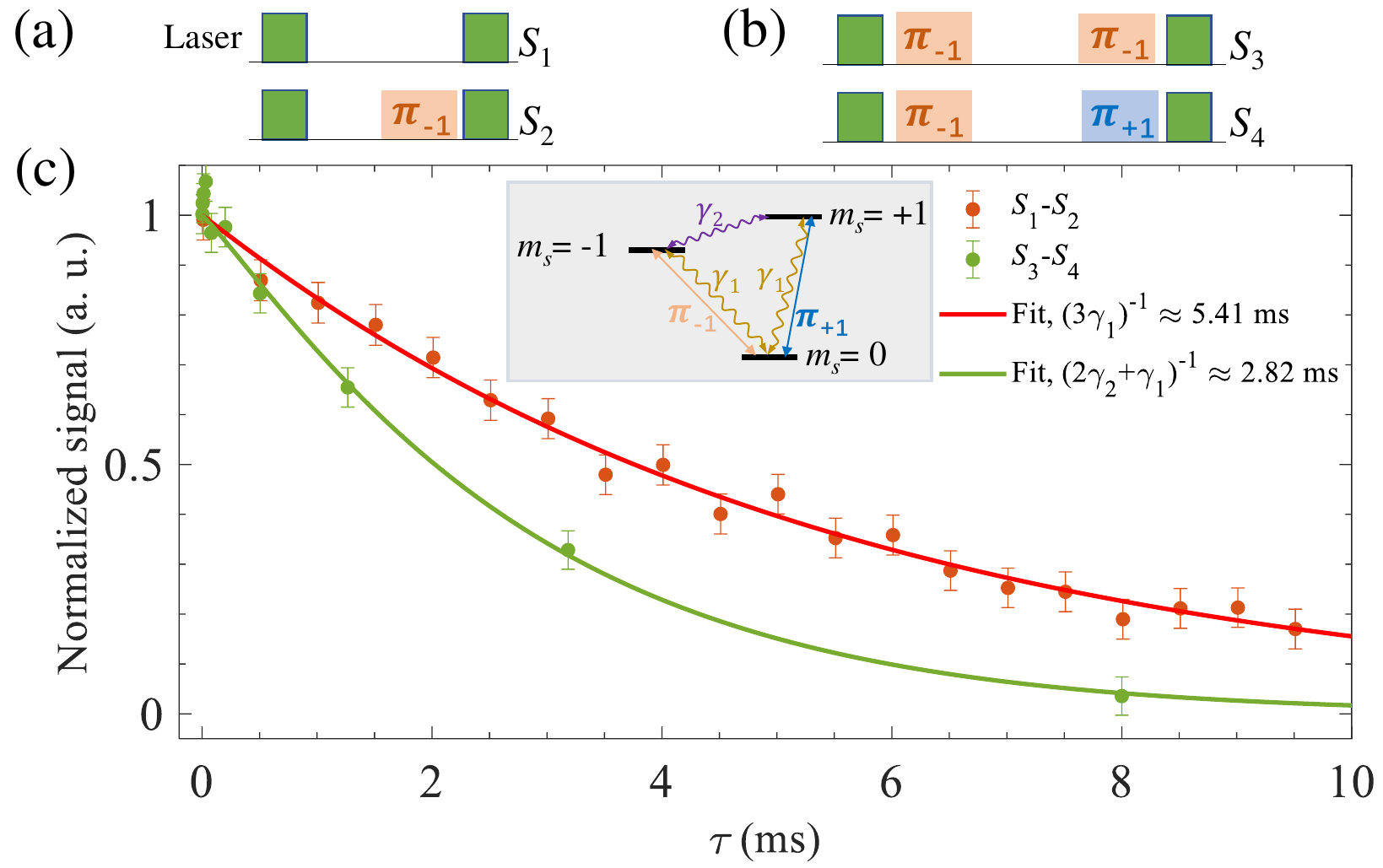}
\centering
\caption{Experimental measurements of the normalized signal difference vs. experimental time for $T_{1}$ estimation. Measuring spin-lattice relaxation time of the difference signal $S_1-S_2$ from an alternating measurement when the system is initialized in the $m_s=0$ state gives $T_1^{(0)}=(3\gamma_1)^{-1}=5.41\pm 0.11$ ms (red dots), where the relation between $T_1^{(0)}$ and $\gamma_1$ follows \cite{myers2017double,cambria2021state}. The difference signal $S_3-S_4$ from an alternating measurement when the system is initialized in the $m_s=-1$ state results in $(2\gamma_2+\gamma_1)^{-1}=2.82\pm 0.23$ ms (green dots) with the difference similar to previously reported results \cite{myers2017double,cambria2021state}. The measurement sequences are shown in the upper sub-panel. Both data sets were fitted to a stretched exponential function. }
	\label{T1_NV}
\end{figure}

In order to characterize spin-lattice relaxation rates, we performed a series of measurements, shown in Fig. \ref{T1_NV}. These characterize the single-quantum relaxation rate $\gamma_1$ between states $|0\rangle\leftrightarrow|\pm 1\rangle$ and the double-quantum relaxation rate $\gamma_2$ between states $|-1\rangle\leftrightarrow|1\rangle$ \cite{myers2017double,cambria2021state}. 
The latter is likely due to  quasilocalized phonons or contributions from higher order terms in the spin-phonon Hamiltonian \cite{cambria2021state}. 
In a first series of measurements we initialize the system in $|0\rangle$ and perform alternating measurements without (with) a $\pi$ pulse on the $|0\rangle\leftrightarrow|-1\rangle$ transition before the readout with the laser pulse. The difference between the resulting signals is normalized and fitted to an exponential decay function, allowing us to obtain $T_1^{(0)}=(3\gamma_1)^{-1}=5.41\pm 0.11$ ms, where the latter is $\pm$ one standard deviation \cite{myers2017double,cambria2021state}.
In a second series of measurements, we initialize the system in $|-1\rangle$ and perform alternating measurements with a $\pi$ pulse on the $|0\rangle\leftrightarrow|-1\rangle$ ($|0\rangle\leftrightarrow|+1\rangle$) transition before the readout with the laser pulse. The difference between the resulting signals is again normalized and fitted to an exponential decay function, allowing us to obtain $(2\gamma_2+\gamma_1)^{-1}=2.82\pm 0.23$ ms  \cite{myers2017double,cambria2021state}. Finally, the obtained ratio $\gamma_2/\gamma_1\approx 1.87$ is close to $2$, in accordance to what has been observed in previous work \cite{cambria2021state}.

\begin{figure}[t]
\centering
\includegraphics[width=1\linewidth]{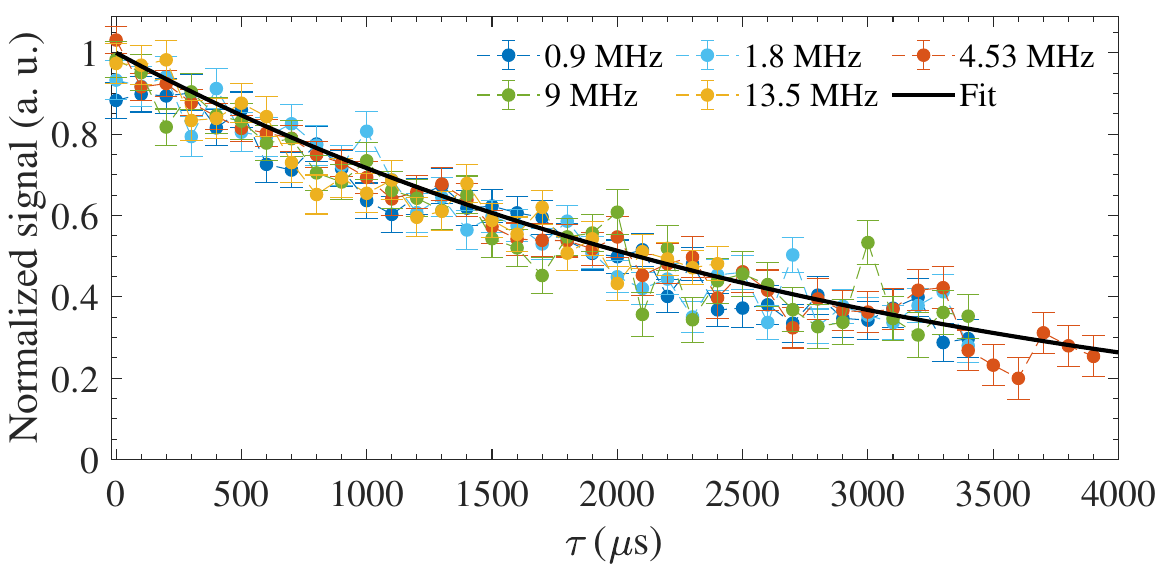}
\centering
\caption{Experimental measurements of the normalized signal difference, obtained from two alternating measurements (see text) vs. experimental time for estimating $T_{1\rho}$ for single drive decoupling. Experiments of several different single drive amplitudes $\Omega_1/2\pi$ were performed. An exponential decay function $\exp(-\tau/3.0~\text{ms})$ (black solid line) is used to fit the data. These measurements show the same $T_{1\rho}$ decay time, of approximately 3.0 ms.}
	\label{T1rho_singledrive}
\end{figure}

In order to assess the limit imposed by relaxation on the coherence time when using decoupling techniques, we conducted an experimental investigation of the dressed qubit relaxation time, $T_{1\rho}$, for a single-drive and for the DD protocols. Measurements of the single-drive $T_{1\rho,single-drive}$ are presented in Fig. \ref{T1rho_singledrive}. The measurements show $T_{1\rho,single-drive}\approx 3$ ms for a wide range of Rabi frequencies. We note that we apply a $\pi_{y}/2$ preparation and $\pi_{\pm y}/2$ final population mapping pulses for this measurement as the eigenstates of the dressed qubit for single-drive decoupling are along the $x$ axis of the Bloch sphere. 

\begin{figure}[t]
\centering
\includegraphics[width=1\linewidth]{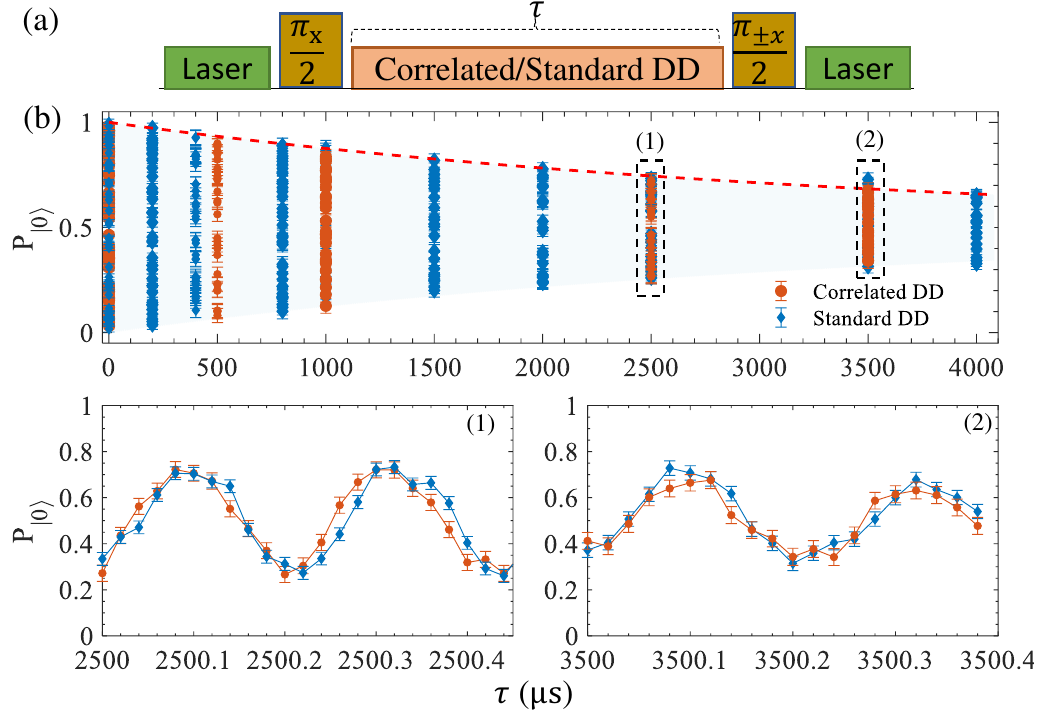}
\centering
\caption{Experimental measurements of the population of state $|0\rangle$ ($P_{|0\rangle}$) vs. experimental time $\tau$ for estimating $T_{1\rho}$ for standard and correlated DD. (a) Scheme of sequence used to measure $T_{1\rho}$. (b) Experimental data for standard and correlated DD, denoted as blue and orange data, respectively. An exponential decay function $1/2+1/2\exp(-\tau/3.5~\text{ms})$ (red dashed line) is used to fit the data envelope. Both DD protocols show the same $T_{1\rho}$ decay time, of approximately 3.5 ms. Zoomed-in data traces at $\tau=$2.5 ms and 3.5 ms are shown in sub-panels (1), (2).}
	\label{T1rho}
\end{figure}

In the case of the DD protocols, the sequence used in the experiment is depicted in Fig. \ref{T1rho}a and includes a $\pi_{x}/2$ pulse that prepares the spin state along the $y$ axis on the Bloch sphere, followed by (standard or correlated) DD. In the second rotating frame, the effective-drive appears as a constant field parallel to the spin's orientation, thereby satisfying the 'spin-locking' condition\,\cite{loretz2013radio}. After the DD interaction for a duration of $\tau$, the qubit state is transferred back to a detectable population using a $\pi_{x}/2$ pulse (or a $\pi_{-x}/2$ pulse for alternating measurements). We perform the readout by a laser pulse, which simultaneously prepares the qubit for the subsequent experiment. The measured signal oscillates with a frequency $\Omega_{1}$, as expected from theory, and decays over a timescale corresponding to the rotating frame relaxation time $T_{1\rho,DD}$. The results, as shown in Fig. \ref{T1rho}, reveal that both standard and correlated DD result in the same relaxation time, approximately $T_{1\rho,DD}\approx3.5\;$ms, which is also expected from theory. We note that for correlated DD, the spin-locked state is slightly shifted from the $y$ axis, due to the detuning of $\widetilde{\Omega}_1$ from $\Omega_1$. However, this difference was insignificant in our measurements. 

\subsection{3e: Estimation of the relaxation-free coherence time}
By neglecting the finite duration time of the XY8 pulses (the duty cycle is $\approx 2.2\%$), the XY8 coherence time is relaxation-limited by $T_{1\rho, single-drive}=3$ ms. In addition to standard relaxation effects, this limit takes into account high-frequency magnetic noise along the z-axis, which contributes to the rotating-frame relaxation and cannot be compensated for by standard control techniques. In the case of ideal $\pi$ pulses the coherence time of the XY8 sequence should be the same as the one of CPMG \cite{carr1954effects,MeiboomGill1960}, with the latter approaching $T_{1\rho,single-drive}$ for small pulse separation \cite{Santyr1988JMR}. Thus, the relaxation-free coherence time, $T_{\varphi,XY8}$, can be estimated by solving $\frac{1}{T_{2,XY8}}=\frac{1}{T_{1\rho,single-drive}}+\frac{1}{T_{\varphi,XY8}}$ \cite{krantz2019quantum,wang2020coherence}, which gives 10.24 ms. Note that this estimation assumes that all signal envelopes can be fitted well by a decaying exponential function, which is indeed the case in our experimental data. To estimate the relexation-free coherence time for the CDD protocols, we conducted simulations of the Lindblad equation  
\begin{equation}
    \partial_t \rho = - i [H,\rho] + \sum_{i=1}^7 \Gamma_i (L_i \rho L_i^\dagger- \frac{1}{2} \{L_i^\dagger L_i,\rho \}),
\end{equation}
in the first interaction picture. The single-quantum jump operators are $L_1=\ket{-1}\bra{0}=L_2^\dagger \And L_3=\ket{0}\bra{1}=L_4^\dagger$, the double-quantum jump operators are $L_5=\ket{-1}\bra{1}=L_6^\dagger$, and the pure-dephasing operator is $L_7=S_z=(\ket{1}\bra{1}-\ket{-1}\bra{-1})$. The single-quantum ($\Gamma_{1,2,3,4}=\gamma_1$) and double-quantum ($\Gamma_{5,6}=\gamma_2$) relaxation rates were set according to the $T_1$ measurements described in the previous section. The pure dephasing rate, $\Gamma_7$, was set at 360 Hz, to reproduce the single-drive rotating-frame relaxation measurement $T_{1\rho,single-drive}$. This leaves no free parameters in the model, and its validity is supported by producing the correct CDD double-rotating-frame relaxation rate $T_{1\rho,CDD}=3.5$ ms. Under these conditions, the simulation gives $T_{2\rho,CDD}^{limit}=3.35$ ms, and thus we can solve $\frac{1}{T_{2\rho,CDD}}=\frac{1}{T_{2\rho,CDD}^{limit}}+\frac{1}{T_{\varphi,CDD}}$ which gives $T_{\varphi,CDD}=17.05$ ms. Thus, $T_{\varphi}$ is longer by $67\%$ compared to XY8. Since the double-rotating-frame relaxation rate $T_{1\rho,CDD}$ in simulation matches the experiment, we conclude that the coherence time $T_{\varphi,CDD}$ is limited by second-order fluctuations of the amplitude noise. Thus, the coherence time can be further improved, to reach the relaxation-induced limit, by optimizing the Rabi frequencies ($\Omega_{1,2}$) and using our protocol as a building block for more complicated schemes such as rotary echos, composite pulses and pulse sequences.

\section{Note 4: Numerical Simulations Details}\label{Section:Numerics}

We perform a numerical simulation where we apply DD in a two-state system, which is subject to environmental noise $\delta(t)$ and correlated amplitude fluctuations of the driving fields. The noise parameters are typical for experiments in NV centers, as described in \cite{CaiNJP2012,AharonNJP2016}. Specifically, the noise $\delta(t)$ is modelled as an Ornstein-Uhlenbeck (OU) process \cite{UhlenbeckRMP1945,GillespieAJP1996,GillespieAJP1996a} with a zero expectation value $\langle \delta(t)\rangle = 0$, correlation function $\langle \delta(t)\delta(t^{\prime})\rangle =(1/2)D\widetilde{\tau}\exp{(-\gamma|t-t^{\prime}|)}$, where $D$ is a diffusion constant and $\widetilde{\tau}=1/\gamma$ is the correlation time of the noise. The OU process is implemented with an exact algorithm \cite{GillespieAJP1996,GillespieAJP1996a}
\begin{equation}\label{Eq:OU_noise}
\delta(t+\Delta t)=\delta(t) e^{-\frac{\Delta t}{\widetilde{\tau}}}+\widetilde{n}\sqrt{\frac{D\widetilde{\tau}}{2}\left(1-e^{-\frac{2\Delta t}{\widetilde{\tau}}}\right)},
\end{equation}
where $\widetilde{n}$ is a unit Gaussian random number. In a first simulation, we assume a correlation time of the noise of $\widetilde{\tau}=25 \mu$s with a diffusion constant $D\approx 4/(T_{2}^{\ast 2} \widetilde{\tau})$, where $T_{2}^{\ast}=3.6 \mu$s, similarly to \cite{CohenFP2017,Genov2019MDD}. The driving fluctuations are also modelled by correlated OU processes with the same correlation time $\tau_{\Omega}=500 \mu$s and a relative amplitude error $\delta_{\Omega}=0.005$ with the corresponding diffusion constant $D_{\Omega_{i}}=2\delta_{\Omega}^2\Omega_{i}^2/\tau_{\Omega},~i=1,2$, following  \cite{CohenFP2017,Genov2019MDD}.

We note that we model the system in the first interaction picture, after applying the first rotating-wave approximation ($\Omega_1,\Omega_2,\widetilde{\Omega}_1\ll\omega_0$), with the Hamiltonian $H_I$ in Eq. \eqref{Eq:H_I}. 
We choose not to work with the second interaction picture Hamiltonian (Eq. \eqref{Eq:HII}), as $\delta(t)$ can often not be neglected. This is especially true when the power of the first driving field is not much greater than zero frequency power spectrum component of the $\delta(t)$ noise. Additionally, we do not apply the second rotating-wave approximation ($\Omega_2\ll\widetilde{\Omega}_1$) in our numerical simulation, which expands significantly the parameter range when it is applicable. 

We proceed by numerically calculating the propagator
\begin{equation}\label{U_numeric}
\widetilde{U}_{I}(t,t_0)=\mathcal{T} \exp{\left(-i\int_{t_0}^{t}H_{I}(t^{\prime})d t^{\prime}\right)}
\end{equation}
for the particular noise realisation of $\delta(t)$, $\epsilon_1(t)$ and $\epsilon_2(t)$ and the chosen DD sequence. We denote the propagator with a tilde to emphasize that it contains noise terms. We use a time-discretization with a time step of $0.1$ ns, which is comparable to the resolution of many standard arbitrary wave-form generators. We note that the OU noise characteristics are not affected by this choice of $\Delta t$, as Eq. \eqref{Eq:OU_noise} is exact.

In order to characterize the performance of our protocols for a quantum memory, we simulate the  fidelity, defined in Eq. \eqref{eq:Fidelity}, which does not depend on the particular initial state, as proposed in \cite{BowdreyPhLett2002} and applied in \cite{Genov2019MDD}. For $\mathcal{M}=\mathbb{1}$, it can be shown to be equivalent to 
\begin{equation}
\widetilde{F}(t)=\frac{1}{3} \sum_{k=x,y,z}\text{Tr}\left( \widetilde{U}_{I}(t,t_0) \rho_{k}(t_0) \widetilde{U}_{I}(t,t_0)^{\dagger} \rho_{k}(t_0)\right)
\end{equation}
for a particular noise realization, where $\rho_{k}(t_0)=(\sigma_{0}+\sigma_{k})/2,~k=x,y,z$ are the density matrices at the initial time $t_0$ corresponding to states along the $x$, $y$, and $z$ axes of the Bloch sphere. 

\begin{figure*}[t!]
	\centering
	\includegraphics[scale=0.65]{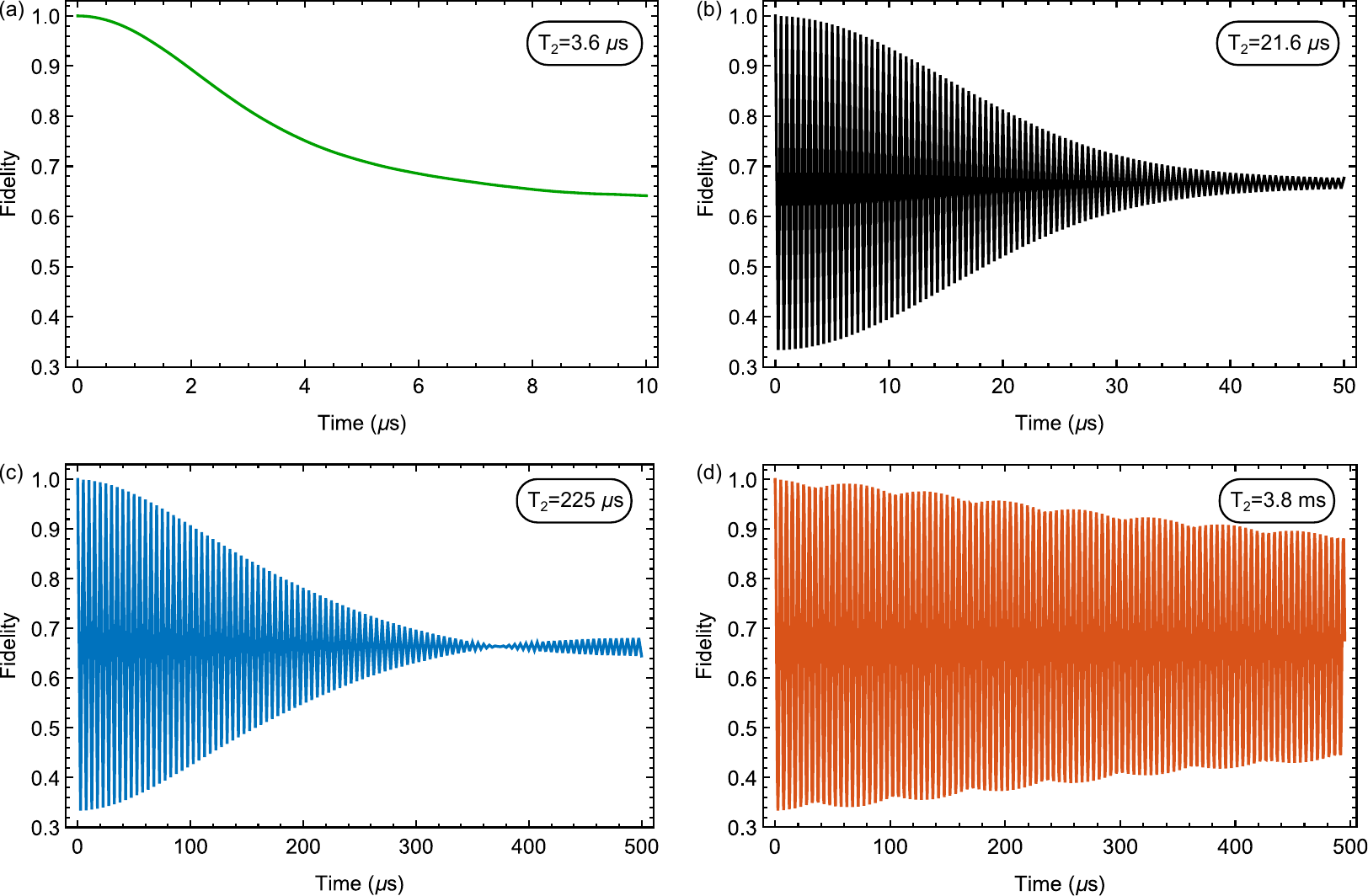}
	\centering
\caption{Simulation of noise protection performance for different sequences: (a) Pure decoherence without any field, (b) Noisy single-drive with $\Omega_1=2\pi~2$ MHz, (c) Standard double-drive with perfectly correlated amplitude noise terms with $\Omega_1=2\pi~2$ MHz, $\Omega_2/\Omega_1=0.1$. The frequency of the second drive is $\widetilde{\Omega}_1=\Omega_1$ , (d) Correlated double-drive with perfectly correlated amplitude noise terms with $\Omega_1=2\pi~2$ MHz, $\Omega_2/\Omega_1=0.1$. The frequency of the second drive is shifted to $\widetilde{\Omega}_1=\Omega_1+\frac{\Omega_2^2}{\Omega_1}$. Under the conditions in the simulation, the correlated double-drive scheme is limited by second-order contributions of the control noise.
}
\label{Fig3:QM_fidelities}
\end{figure*}

Next, we make use of the numerically calculated propagator $\widetilde{U}_{I}(t,t_0)$ in Eq. \eqref{U_numeric} to obtain the time-dependence of the density matrices for the particular noise realization for the three particular initial states
\begin{equation}
\rho_{k}(t)=\widetilde{U}_{I}(t,t_0)\rho_{k}(t_0)\widetilde{U}_{I}(t,t_0)^\dagger,~k=x,y,z
\end{equation}
Then, we calculate the expected density matrix $\overline{\rho}_{k}(t)$ for each of these three initial states by averaging the density matrices  $\rho_{k}(t)$ for all noise realizations. This allows us to calculate the expected fidelity at time $t$
\begin{equation}
F(t)=\frac{1}{3}\sum_{k=x,y,z} \text{Tr} \left(\overline{\rho}_{k}(t)\rho_{k}(t_0) \right).
\end{equation}

The expected average fidelity is calculated by performing the simulation $2500$ times for different noise realizations.
Our simulation agrees well with previous numerical results for NV centers \cite{AharonNJP2016}. For example, Fig. \ref{Fig3:QM_fidelities}(a) shows that our simulation exhibits the expected decoherence time of $T_{2}^{\ast}\approx\,3.6 \mu$s without control fields. We note that we define the $T_{2}^{\ast}$ (and $T_{2}$ for the other protocols) of the average fidelity as the time it takes it to drop to from $1$ to $\approx 0.79$, which corresponds to a $1/e$ drop in the difference from $1$ to the limit of $0.67$ (this limit is conservative as it assumes negligible decay for one of the initial states because the simulation does not consider population relaxation) \cite{Genov2019MDD}. The coherence time increases to tens of microseconds when decoupling with a single drive (Fig. \ref{Fig3:QM_fidelities}(b)). Standard DD prolongs the coherence time further to $225~\mu$s with amplitude noise being the main reason for the remaining decay, as shown in Fig. \ref{Fig3:QM_fidelities}(c). 
Correlated DD achieves a coherence time of $3.8$ ms, an improvement of twenty times over standard DD (see Fig. \ref{Fig3:QM_fidelities}(d)), which is of the same order as in the experiment, described in the main text. The simulation does not account for relaxation processes and uses lower Rabi frequencies than the experiment, resulting in longer coherence times. 
We note that the population relaxation time of an NV center can reach
up to $6$ ms \cite{Bar-GillNatComm2013},
so the fidelity at long storage times with correlated DD can be affected by the population relaxation time of the system  (not taken into account in the simulations) in this particular implementation.

In addition to the average fidelity simulation, we also perform simulations of the fidelities for different initial states
\begin{equation}
F_k(t)=\text{Tr} \left(\overline{\rho}_{k}(t)\rho_{k}(t_0) \right),~k=x,y,z,
\end{equation}
as they correspond directly to experimental results. For example, a simulation of $F_x(t)$ corresponds to the fidelity of the standard and correlated DD protocols in the experiment in Fig. 1
(c) in the main text and allows for calculating $T_{2\rho}$. This is the case because the applied initial $\pi_{y}/2$ pulse before the DD protocol ideally creates a coherent superposition state along the $x$ axis of the Bloch sphere, which makes the simulation of $F_x(t)$ relevant for the experiment. Similarly, a simulation of $F_y(t)$  corresponds to the experiment in Fig. \ref{T1rho} because of the applied initial $\pi_{x}/2$ pulse and allows for the calculation of $T_{1\rho}$ for the standard and correlated DD protocols. We note that we define the $T_{2\rho}$ of the fidelities $F_k(t),~k=x,y,z$ as the time it takes them to drop to from $1$ to $\approx 0.684$, which corresponds to a $1/e$ drop in the difference from $1$ to the quantum limit of $0.5$. 

Next, we perform simulations of $F_k(t)$ and calculate the respective coherence times for single drive decoupling, standard and correlated DD for different noise spectra of the amplitude noise. The simulation results are summarized in Fig. 3 
in the main text, and we provide additional details here and in Table \ref{Table:coherence_times}. In all simulations, we assume the same parameters for the $\delta(t)$ noise as above, i.e., a correlation time of the noise of $\widetilde{\tau}=25 \mu$s with a diffusion constant $c\approx 4/(T_{2}^{\ast 2} \widetilde{\tau})$, where $T_{2}^{\ast}=3.6\,\mu$s. The Rabi frequencies of the two driving fields are also the same as in the simulation in Fig. \ref{Fig3:QM_fidelities}, i.e., $\Omega_1=2\pi\,2$ MHz, $\Omega_2=2\pi\,0.2$ MHz with $\Omega_2/\Omega_1=0.1$.

\begin{table}[t]
\caption{Amplitude noise characteristics: correlation time $\tau_{\Omega}$, the corresponding $\delta_{\Omega}$, and the respective coherence times $T_{2\rho}$ for single drive decoupling, standard DD (SDD) and correlated DD (CDD, $c\to 1+\frac{1}{4}$
).
}
\begin{tabular}{l l l r r} 
\hline 
$\tau_{\Omega}~(\mu s)$~~~& $\delta_{\Omega}~(\%)$~~~& Protocol~~&~~~~$T_{2\rho}~(\mu s)$& ~$T_{2\rho}/T_{2\rho,\text{single}}$ \\ 
\hline 
0.5 & 2.4 & Single drive & 21 & 1 \\
0.5 & 2.4 & SDD & 67 & 3.2 \\
0.5 & 2.4 & CDD & 68 & 3.2 \\
\hline 
5 & 0.85 & Single drive & 22 & 1 \\
5 & 0.85 & SDD & 526 & 24 \\
5 & 0.85 & CDD & 1126 & 52 \\
\hline 
25 & 0.58 & Single drive & 22 & 1 \\
25 & 0.58 & SDD & 470 & 22 \\
25 & 0.58 & CDD & 4100 & 190 \\
\hline 
50 & 0.54 & Single drive & 22 & 1 \\
50 & 0.54 & SDD & 325 & 15 \\
50 & 0.54 & CDD & 3968 & 182 \\
\hline 
100 & 0.52 & Single drive & 22 & 1 \\
100 & 0.52 & SDD & 247 & 11 \\
100 & 0.52 & CDD & 2816 & 131 \\
\hline 
500 & 0.51 & Single drive & 22 & 1 \\
500 & 0.51 & SDD & 210 & 10 \\
500 & 0.51 & CDD & 1816 & 83 \\
\hline 
\end{tabular}
\label{Table:coherence_times} 
\end{table}

We perform a set of simulations where we vary the correlation time of the amplitude noise $\tau_{\Omega}$ in the range between $0.5$ and $500\,\mu$s and the relative amplitude error standard deviation $\delta_{\Omega}$ between $0.0051$ (or equivalently $0.51\%$) and $0.024$ (or equivalently $2.4\%$). The value of the correlation time determines the width of the respective amplitude noise spectrum $\approx 1/\tau_{\Omega}$ while the corresponding value of $\delta_{\Omega}$ has been chosen such that the coherence time of the single drive decoupling protocol is approximately the same for all probed noise spectra, thus, the self-averaging effect of using faster noise is compensated for. A summary of the coherence times  for the different protocols and the corresponding noise spectra characteristics is given in Table \ref{Table:coherence_times}. We note that the coherence time $T_{2\rho}$ had been calculated for $F_y(t)$ for single-drive decoupling and $F_x(t)$ for standard double drive (SDD) and correlated double drive (CDD) as these are the fidelities of the fast decaying states for the respective protocol. The simulation results show that correlated DD outperforms standard DD for all correlation times (and corresponding noise spectra) and the improvement is greatest when the $\tau_{\Omega} \in (2\pi/\Omega_1,2\pi/\Omega_2)$, where $\Omega_{1,2}$ are in angular frequency units. 

\section{Note 5: Quantum Sensing}
\section{5a: Sensing Schemes with Correlated Double Drive}\label{Section:Quantum_Sensing_SI}

One advantage of the prolongation of coherence times is the enhancement of the sensitivity of quantum sensors \cite{Degen2017RMP}. Specifically, standard DD protocols were used for AC magnetic field sensing with NV centers in the high-frequency (GHz) \cite{StarkNatComm2017} and low frequency (Sub-MHz) \cite{https://doi.org/10.48550/arxiv.2207.06611} domains. To demonstrate the use of the correlated DD protocol for enhanced quantum sensing, we consider the high-frequency signal $H^{g}=g_0\sin(\omega_g t)\sigma_x$, which is added to the lab-frame Hamiltonian in Eq. \eqref{Eq:lab_frame_H}.
Following the analysis of the previous section, we move the signal to the second interaction picture and transform it once more using $U_0^{(3)}$, to obtain
\begin{equation}\label{Eq:H_Signal}
\begin{aligned}
    &H_{IIS}+H^{g}_{IIS}=
    -\frac{\Omega_e}{2}\sigma_x+\\&
    \frac{g'}{2}
    (\sin(\Omega_e t+\varphi)\sigma_y+\cos(\Omega_e t+\varphi)\sigma_z)
\end{aligned}
\end{equation}
where we have omitted the noise terms for simplicity and assumed that the signal frequency obeys one of the resonance conditions described shortly. The effective signal amplitude is $g'=\alpha g_0$, where $\alpha=\alpha_{DD}\widetilde{\alpha}$ is the signal attenuation factor that has two contributions. $\alpha_{DD}$ - existing for both DD schemes and $\widetilde{\alpha}$ - resulting from the frequency shift necessary for the correlated DD protocol. $\varphi$ is a constant phase of little importance for the following discussion. All three parameters depend on the choice of $\omega_g$. The Hamiltonian in Eq. \eqref{Eq:H_Signal} allows sensing of the signal $H^{g}$ by meeting a resonance condition for $\omega_g$ and detecting Rabi oscillations in the second dressed basis.

Past experiments \cite{StarkNatComm2017} using the standard DD protocol ($\widetilde{\Omega}_1=\Omega_1$) demonstrated quantum sensing of a signal with frequency $\omega_g=\omega_0-\widetilde{\Omega}_1-\Omega_e$, in which case the total signal attenuation factor was $\alpha=\alpha_{DD}=\frac{1}{4}$ ($\widetilde{\alpha}=1$). The generalized version of this sensing experiment, with the correlated-noise-shift, results in the parameters reading $\alpha_{DD}=\frac{1}{4}$, $\widetilde{\alpha}=\frac{(\Omega_1-\widetilde{\Omega}_1)+\Omega_e}{\Omega_e}$ and $\varphi=\frac{\pi}{2}$. We refer to this setting as the ``high-attenuation'' sensing scheme. Note that in the standard parameters regime, e.g. $\frac{\Omega_2}{\Omega_1}\sim0.2$ and $\widetilde{\Omega}_1$ set according to Eq. \eqref{Eq:Shift_Corr_1}, we have $\widetilde{\alpha}\approx\frac{4}{5}$ and thus $\alpha\approx\frac{1}{5}$. We further note that in this work \cite{StarkNatComm2017}, it was demonstrated that the sensed field can act as a third decoupling drive, further protecting the sensor coherence. In real sensing scenarios, however, the sensed field's amplitude might be weak, rendering this coherence protection effect negligible.

We note, however, that sensing using a different resonance condition results in smaller attenuation factors, which in turn improves the sensitivity. The ``low-attenuation'' sensing scheme, therefore, entails setting  $\omega_g=\omega_0-\Omega_e$, and the parameters now read $\alpha_{DD}=\frac{1}{2}$, $\widetilde{\alpha}= \frac{-\Omega_2}{\Omega_e}$ and $\varphi=0$. Thus, the low-attenuation sensing scheme is better due to two reasons. First, $\alpha_{DD}$ is improved by a factor 2, rendering the DD schemes' signal attenuation equal to single-drive schemes \cite{StarkNatComm2017}. Second, for $\frac{\Omega_2}{\Omega_1}=0.2$ and $\widetilde{\Omega}_1$ set according to Eq. \eqref{Eq:Shift_Corr_1} we have $\widetilde{\alpha}\sim1$, so the frequency shift results in negligible signal attenuation. The experimental results comparing the low-attenuation scheme for both DD protocols are shown in the main text. We note that to investigate only the decoupling effect of the control method, we generated a signal with a relatively weak amplitude ($g_0$ of a few kHz). 
Importantly, since the attenuation factors modify the effective coupling to external fields, this improvement confers an advantage for quantum computing schemes which rely on dressed qubits \cite{CaiNJP2012}. 

\begin{figure}[t]
\centering
\includegraphics[width=1\linewidth]{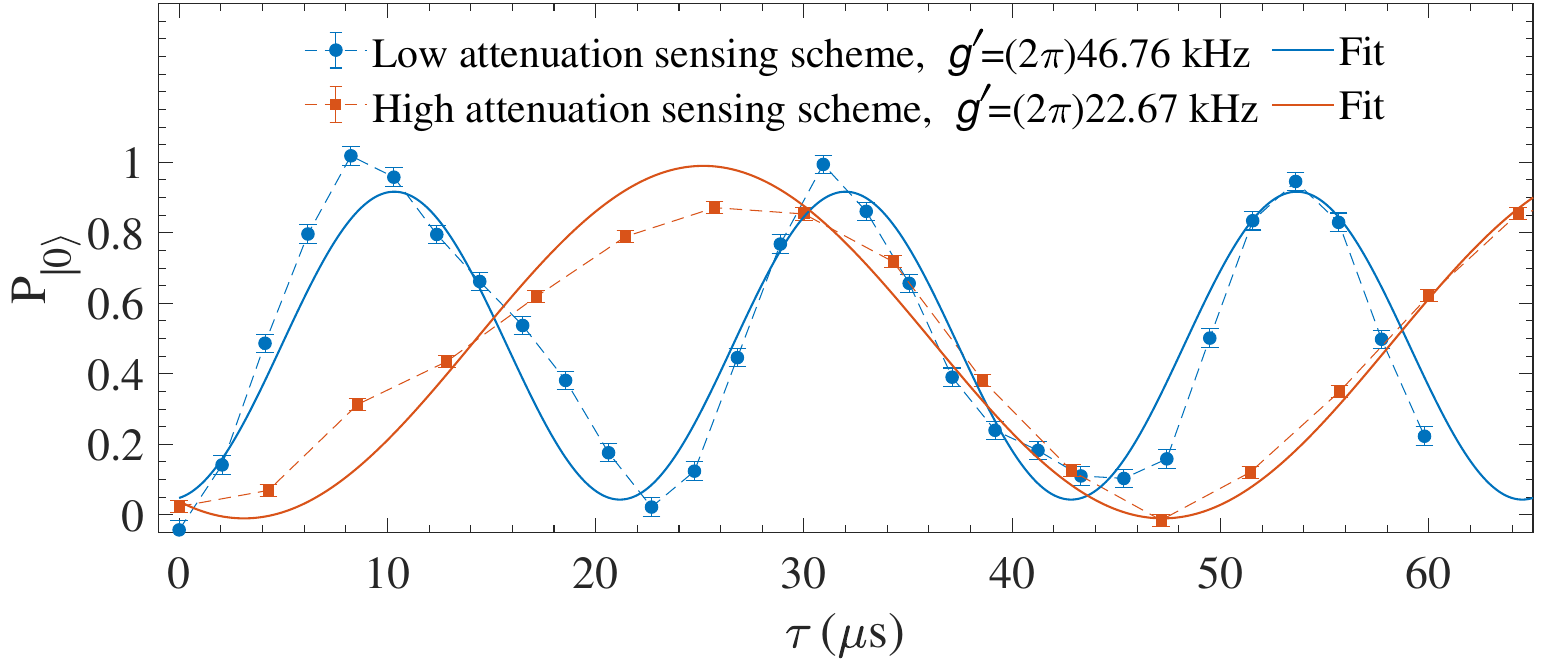}
\centering
\caption{Quantum sensing with improved coupling to external fields. The low and high attenuation schemes are compared by sensing an external signal of strength $g\approx(2\pi)94$ kHz. An effective signal strength $g'=(2\pi)46.76\;\text{kHz}\approx g/2$ is recorded with the low attenuation sensing scheme (blue circles), and $g'=(2\pi)22.67\;\text{kHz}\approx g/4$ with the high attenuation sensing scheme (orange square). The attenuation factors match the theoretical prediction.}
	\label{sensing_scheme_results}
\end{figure}

In a third series of experiments, we compare the low-attenuation sensing scheme we proposed to the previously demonstrated high-attenuation scheme. We generated a signal $g(t)$ with frequency $\omega_g$ set by their respective resonance conditions.
The signal was generated from the same AWG channel which also generates the measurement sequence (DD and $\pi/2$ pulses). Experimentally, we first demonstrated the low-attenuation sensing scheme and set $\Omega_{1}=(2\pi)4.666$ MHz, $\Omega_{2}=(2\pi)0.913\ \text{MHz}\approx\frac{1}{5}\Omega_1$ and $\widetilde{\Omega}_1$ according to Eq. \eqref{Eq:Shift_Corr_1}. The measurement sequence was the same as the one used to measure $T_{1\rho}$ (see Fig. \ref{T1rho}a). To acquire data without the $\Omega_{1}$ oscillation, we sample the system at specific durations that are multiples of $\tau_{\Omega_1}=2\pi/\Omega_{1}$. The measurement results are shown in Fig. \ref{sensing_scheme_results}, where an oscillation frequency $g^{\prime}=(2\pi)46.76\pm0.59\;\text{kHz}\approx g_0/2$ is recorded for the low attenuation sensing scheme. For comparison, we also performed the standard high-attenuation scheme, and an oscillation frequency $g^{\prime}=(2\pi)22.67\pm0.60\;\text{kHz}\approx g_0/4$ is recorded, as expected from theory.

\section{5b: Sensitivity estimation}\label{Section:Sensitivity}

In conventional optical readout of single NV experiments, contrast is typically low, and the number of photons collected per measurement is limited (much less than 1, e.g., $0.06-0.1$), so photon shot-noise becomes the dominant noise source that affects magnetic field sensitivity \cite{barry2020sensitivity, rondin2014magnetometry, taylor2008high}. Therefore, we estimate the sensitivity of the correlated DD for quantum sensing for experimental parameters in the photon shot-noise limited regime.

In both standard and correlated DD sensing experiments (illustrated in Fig. 4 
in the main text), we repeat the measurements $N_{\text{sweep}}$ times, and record the normalized NV photoluminescence (PL) signal $S$ as:
\begin{align}\label{signal_sensing}
\footnotesize
S&= \frac{a+b}{2} - \frac{(a-b)\Gamma(\tau)}{2}\cos( g^{\prime}\tau)\notag\\
&= \frac{a+b}{2} - \frac{(a-b)\Gamma(\tau)}{2}\cos(\alpha \gamma_{\text{NV}}B\tau),
\end{align}
where $a$ and $b$ denote the normalized signal for bright and dark states, respectively; $\Gamma(\tau)$ is a decay function, which characterizes the loss of contrast due to decoherence during the DD sequence. The total attenuation factor, $\alpha=\alpha_{DD}\widetilde{\alpha}$, characterizes the effective coupling strength of the sensed field $g^{\prime}=\alpha g=\alpha\gamma_{\text{NV}} B$ with $\gamma_{\text{NV}}=(2\pi)$28 Hz/nT -- the gyromagnetic ratio of the NV electron spin, $B$ is the magnetic field signal amplitude, and $\tau$ is the sensing time. Assuming that $N_{\text{ph}}$ represents the average number of photons collected per measurement run, the expected total number of photons after $N_{\text{sweep}}$ repetitions is $N_0=aN_{\text{sweep}} N_{\text{ph}}$ for the bright state $|0\rangle$, and $N_1=bN_{\text{sweep}} N_{\text{ph}}$ for the dark state $|1\rangle$. Then we use the formula for the PL signal $S$ to calculate the expected total number of collected photons per data point for all measurements as: 
 \begin{equation}\label{photon_signal}
     N_{S}=S N_{\text{sweep}} N_{\text{ph}} .
 \end{equation}

The uncertainty $\delta N_{S}$ in the resulting signal $N_{S}$ is related to the corresponding uncertainty $\delta B$ in the measured field $B$ by the slope:
\begin{equation}
\delta N_{S} = \frac{\partial N_{S}}{\partial B} \delta B ,
\end{equation}
The signal is most sensitive to small changes in the magnetic field at the point of maximum slope, i.e., when  $\alpha \gamma_{\text{NV}}B\tau\approx \pi/2$ \cite{Degen2017RMP}: 
\begin{equation}
\small
 \begin{aligned}
   \max &\left| \frac{\partial N_{S}}{\partial B}\right| = \\& \max\left|\frac{(a-b)\Gamma(\tau)}{2} (N_{\text{sweep}} N_{\text{ph}}\alpha \gamma_{\text{NV}}\tau) \sin(\alpha \gamma_{\text{NV}}B\tau)\right|  \\
   &=\frac{(a-b)\Gamma(\tau)}{2} N_{\text{sweep}} N_{\text{ph}} \alpha \gamma_{\text{NV}}\tau \\
   &=\frac{C(\tau)}{2} N_{\text{sweep}} N_{\text{ph}}\alpha \gamma_{\text{NV}}\tau , 
 \end{aligned}
\end{equation}
where $C(\tau)=(a-b)\Gamma(\tau)$ is the contrast of the signal $S$ at the particular interaction time $\tau$, taking into account decoherence. Since photon shot-noise arising from the Poissonian distribution of the collected photons usually has the highest noise contribution \cite{taylor2008high,barry2020sensitivity,steiner2010universal}, we approximate the signal uncertainty by $\delta N_{S}\approx \sqrt{N_{\text{sweep}} N_{\text{ph}}}$, where $N_{\text{sweep}} N_{\text{ph}}$ is the total number of collected photons. The minimum resolvable magnetic field $\delta B_{\text{min}}$ is therefore given by:
\begin{equation}
\begin{aligned}
    \delta B_{\text{min}} = \frac{\delta N_{S}}{\max|\frac{\partial N_{S}}{\partial B}|} &= \frac{2\sqrt{N_{\text{sweep}} N_{\text{ph}}}}{C(\tau) N_{\text{sweep}} N_{\text{ph}}\alpha \gamma_{\text{NV}}\tau} \\
    &= \frac{2}{\gamma_{\text{NV}} \alpha C(\tau) \tau \sqrt{N_{\text{sweep}} N_{\text{ph}}} }  \\
    &= \frac{2\sqrt{\tau + t_{\text{r}}}}{\gamma_{\text{NV}} \alpha C(\tau) \tau\sqrt{t N_{\text{ph}}} }  ,
\end{aligned}  
\end{equation}
where $t_{\text{r}}$ is the sequence dead time including laser readout time ($\sim3\;\mu$s), idle time ($\sim1.5\;\mu$s), two $\pi/2$ pulses duration ($\sim30\;$ns) with $t=(\tau+t_{\text{r}}) N_{\text{sweep}}$ the total measurement time \cite{barry2020sensitivity}. The sensitivity $\eta(\tau)$ for a repetitive measurement at $\tau$ after time $t$ is given by the following relation:
\begin{equation}
    \eta(\tau)=\delta B_{\text{min}}\sqrt{t}\approx \frac{2}{\gamma_{\text{NV}}\alpha C(\tau)}\frac{1}{\sqrt{N_{\text{ph}}\tau}},
\end{equation}
where we used that $\tau\gg t_{\text{r}},~t_{\text{r}}\approx5\;\mu$s, so $\sqrt{\tau + t_{\text{r}}}\approx\sqrt{\tau}$. Note that increasing $\tau$ reduces the contrast $C(\tau)=(a-b)\Gamma(\tau)$ due to the NV spin decoherence. The optimal sensitivity is obtained by maximizing $\Gamma(\tau)\sqrt{\tau}$ and depends on the decoherence function $\Gamma(\tau)=\exp{\{-\left(\tau/T_{2\rho}\right)^{p}\}}$ with $p$ depending on the noise spectrum and the decoupling protocol. Typically, a simple exponential decay with $p=1$ is assumed and the optimal sensitivity is
achieved when $\tau\approx T_{2\rho}/2$ \cite{taylor2008high,barry2020sensitivity,Degen2017RMP}. 

\begin{figure}[t]
\centering
\includegraphics[width=1\linewidth]{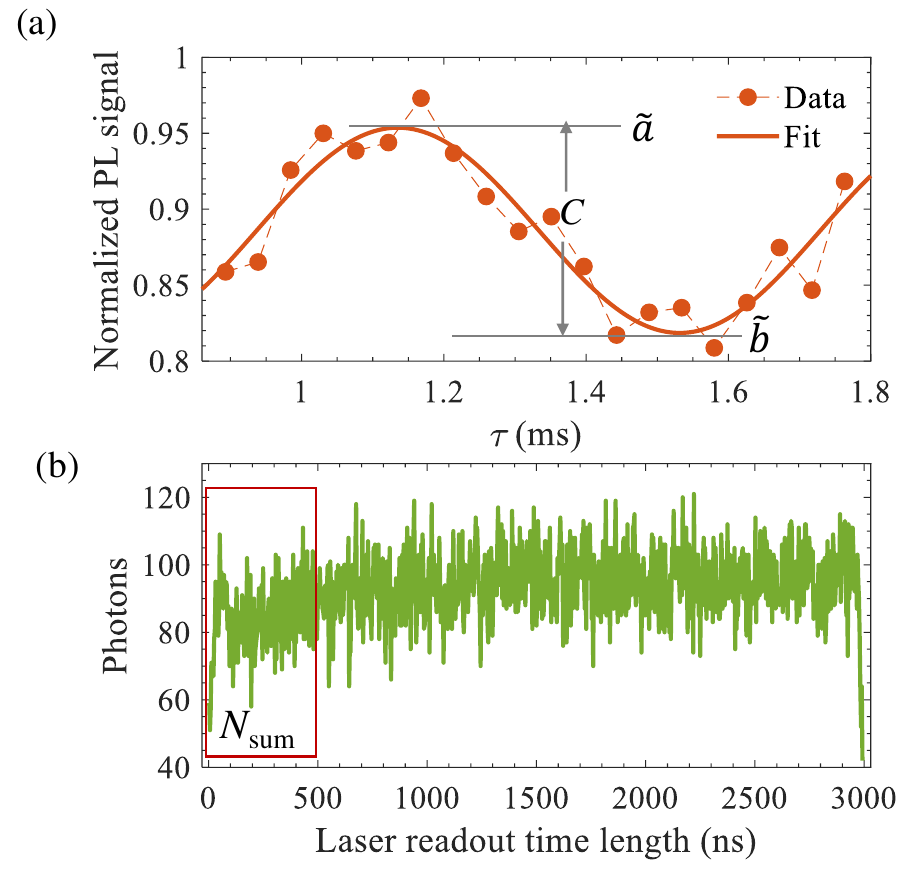}
\centering
\caption{(a) Experimental data of normalized signal $S$ when $\tau\approx 1.3\;$ms in a correlated DD sensing measurements which is shown in Fig. 4 
in the main text. The fitting gives the normalized signal value for bright states $\tilde{a}=a\Gamma(\tau=1.1\;\text{ms})$=0.948, and $\tilde{b}=b\Gamma(\tau=1.5\;\text{ms})$=0.823 for dark states. The contrast is then $C(\tau)=0.125$. (b) Photon count trace accumulated after repeating the measurements and readout for $N_{\text{sweep}}\approx9\times10^4$ times at data point $\tau=1.68\;$ms. Total photons $N_{\text{sum}}$ collected in the first 500 ns(red rectangular area) was used to determine the spin states, here $N_{\text{sum}}\approx1.3\times10^4$. Hence, the average photon number $N_{ph}$ collected in one measurement run is $N_{\text{ph}}=N_{\text{sum}}/N_{\text{sweep}}\approx0.15$. }
	\label{sensitivity_parameter}
\end{figure}

In order to evaluate the experimental photon-shot noise limited sensitivity for the low-attenuation sensing CDD protocol, we use the following experimental parameters. First, the decoherence time is estimated $T_{2\rho}\approx 1.68\;$ms, while $a=1.02$, $b=0.78$ are the normalized signal values for the bright and dark states, respectively, obtained from a Rabi oscillation measurement for very short $\tau$. The expected contrast at time $\tau=T_{2\rho}/2$ is then $C(\tau=T_{2\rho}/2)=(a-b)\Gamma(T_{2\rho}/2)\approx\,0.146$. The measurements were repeated $N_{\text{sweep}}\approx9\times10^4$ times for each experimental point, and the average number of photons collected per measurement run has been estimated $N_{\text{ph}}=N_{\text{sum}}/N_{\text{sweep}}\approx0.15$. Hence, we estimate a photon-shot noise limited sensitivity of
\begin{align}\label{Eq:sensitivity_estimation_1}
    \eta(T_{2\rho}/2)= &\frac{2}{2\pi\times 28\;\text{Hz/nT} \times0.5 \times0.146} \\ 
    &\times \frac{1}{\sqrt{0.15\times 1.682\;\text{ms}/2}}
    \approx\,13.9\;\text{nT/}\sqrt{\text{Hz}}.\notag
\end{align}

We note that the assumption for a simple exponential decay is not always valid and $p$ might differ from one. Therefore, we also evaluated the photon-shot noise limited sensitivity for an example time $\tau\approx 1.3\;\text{ms}$, which lies between $T_{2\rho}/2$ and $T_{2\rho}$, where the curve of our normalized sensed signal crosses the middle value between $a$ and $b$ and the sensitivity is expected optimal. Fig. (\ref{sensitivity_parameter}.a) shows the experiment data of the normalized signal $S$ in the correlated DD sensing measurements, which are also shown in Fig. 4 in the main text. 
The normalized signal values are $\widetilde{a}=a\Gamma(\tau=1.1\;\text{ms})$=0.948, and  $\widetilde{b}=b\Gamma(\tau=1.5\;\text{ms})$=0.823 for the bright and dark states, respectively. The contrast is then $C(\tau\approx 1.3\;\text{ms})=(a-b)\Gamma(\tau)=\widetilde{a}-\widetilde{b}\approx\,0.125$. For each data point in Fig. (\ref{sensitivity_parameter}.a), the measurements were repeated for $N_{\text{sweep}}\approx9\times10^4$ times. 
The accumulated photon count trace at $\tau=$1.68 ms data point is shown in Fig. (\ref{sensitivity_parameter}.b). The total  collected photon counts $N_{\text{sum}}\approx1.3\times10^4$ during the initial 500 ns time window (red rectangular area in Fig. (\ref{sensitivity_parameter}.b)) are used to determine the spin states. This gives the average photon number collected per measurement run of $N_{\text{ph}}=N_{\text{sum}}/N_{\text{sweep}}\approx0.15$. Hence, we conclude that the photon shot noise limited sensitivity is
\begin{equation}
\begin{aligned}\label{Eq:sensitivity_estimation_2}
    \eta(\tau\approx 1.3\;\text{ms})= &\frac{2}{2\pi\times 28\;\text{Hz/nT} \times0.5 \times0.125} \\ 
    &\times \frac{1}{\sqrt{0.15\times 1.3\;\text{ms}}}\approx 13\;\text{nT/}\sqrt{\text{Hz}},
\end{aligned}
\end{equation}
which is slighly better than the previously estimated value of $\approx 13.9\;\text{nT/}\sqrt{\text{Hz}}$ in Eq. \eqref{Eq:sensitivity_estimation_1}. Given the results of the two approaches, we assume that the CDD sensing experiment demonstrates a photon-shot noise limited sensitivity of $\eta\approx 13\;\text{nT/}\sqrt{\text{Hz}}$, which we give in the main text. 

In order to compare performance, we also estimate the sensitivity of the standard DD protocol. We apply the first approach we used for assessing the CDD photon shot noise-limited sensitivity due to its simplicity. The main difference is that $T_{2\rho,\text{SDD}}\approx 494\,\mu$s is more than a factor of three shorter than for CDD. We use that the contrast $C(\tau=T_{2\rho,\text{SDD}}/2)=(a-b)\Gamma(\tau)=0.146$ is expected to be the same at half the coherence time for both CDD and SDD, as the decay is assumed to have the same simple exponential shape for both protocols. In addition, standard DD has been very sensitive to amplitude noise, which required post-selection of results. Specifically, the amplitudes of the Rabi frequencies of the two driving fields had to be remeasured, and the corresponding parameters adjusted practically every measurement point, resulting in an overhead time and an increase in the total measurement time by a factor of approximately three for the same number of sweeps and measurement points. In comparison, such post-selection and overhead time were not necessary for the CDD protocol because of its robustness. The SDD sensitivity then takes the form \cite{taylor2008high,barry2020sensitivity,Degen2017RMP}
\begin{equation}
    \eta_{\text{SDD}}(\tau)=\frac{2}{\gamma_{\text{NV}}\alpha C(\tau)}\frac{1}{\sqrt{N_{\text{ph}}\tau}}\sqrt{\frac{\tau+t_{\text{overhead}}}{\tau}},
\end{equation}
with its optimum value obtained for $\tau=T_{2\rho,\text{SDD}}/2$: 
\begin{align}\label{Eq:sensitivity_estimation_SDD}
    \eta_{\text{SDD}}&\left(\frac{T_{2\rho,\text{SDD}}}{2}\right)= \frac{2}{2\pi\times 28\;\text{Hz/nT} \times0.5 \times0.146} \\ 
    &\times \frac{\sqrt{3}}{\sqrt{0.15\times 0.494\;\text{ms}/2}}
    \approx\,45.8\;\text{nT/}\sqrt{\text{Hz}}.\notag
\end{align}
Thus, the sensitivity of SDD is about 3.3 times worse that the CDD sensitivity, mainly due to the shorter coherence time and much larger post-processing overhead.  

Finally, we note that the enhanced photon collection efficiency also contributes to the improved sensitivity in comparison to previous experiments. The single NV in our diamond SIL sample showed a saturated photon flux of more than 1.3$\times10^3$ kHz. This is one order of magnitude improvement in count-rate compared with traditional sample\cite{hadden2010strongly,siyushev2010monolithic}. We finally note that further optimization of the sensitivity is definitely possible, for example, by optimizing the Rabi frequencies for best coherence time \cite{wang2020coherence}, but this goes beyond the scope of this work. 

\section{Note 6: Robust Coherent Control}\label{Section:RobustCoherentControl}

Even without the need to decouple quantum systems from environmental noise, control fields are used to precisely manipulate quantum states, e.g., by applying pulses. Noise in the driving fields then translates into errors in the desired operations, which in turn leads to gate fidelity losses in the context of quantum computation or reduced metrological sensitivity in the case of quantum sensors. When such noise limits the robustness of control pulses, a destructive interference-based protocol can provide the necessary error compensation. This is demonstrated for population transfer using $\pi$-pulses with the correlated DD protocol in Fig. 5 in the main text. 
We compare three different schemes - a conventional $\pi$-pulse, standard DD and correlated DD, under conditions of quasi-static noise - the dynamics is propagated using the Hamiltonian in Eq. (2) in the main text 
under the assumption of fully-correlated, time-independent, random noise (the value in the x-axis). Fig. 5 in the main text 
demonstrates that the correlated DD protocol significantly outperforms the other methods. The importance of the time overhead factor required for the implementation of different control strategies, compared to the standard pulse, depends on the full physics of the system in question (e.g. its $T_1$ time). We note that in the example in Fig. 5 in the main text 
correlated DD is faster than standard DD, although slower than the standard pulse.  Finally, we emphasize that composite pulses and other refocusing based methods can be combined with destructive interference based methods to provide superior performance.

\paragraph{Commensurability conditions.---}

Achieving high-fidelity operations usually requires that in the absence of noise, the desired operation is performed exactly.
For the standard DD protocol, a straightforward way to achieve this is by choosing $\frac{\Omega_2}{\Omega_1} = \frac{1}{4}$, such that for a total pulse time of $\frac{\pi}{\Omega_2}=\frac{4\pi}{\Omega_1}$ the second interaction picture (given by $H_0^{(2)}=\Omega_1 \sigma_{x}/2$) coincides with the first, and the effective Hamiltonian $H_{II}$ results in a unitary operation equal to that of an exact standard $\pi$ pulse. 
For the correalted DD protocol, the same consideration results in a slightly modified matching condition $\frac{\Omega_2}{\Omega_1} = \frac{1}{\sqrt{15}}\approx0.258)$ and total pulse time $\frac{\frac{15}{4} \pi}{\Omega_1}$.

\bibliography{references}

\end{document}